\documentclass[twocolumn]{aastex631}
\usepackage{gensymb}
\usepackage{hyperref}
\usepackage{amsmath}
\usepackage{amssymb}
\usepackage{latexsym}
\usepackage{graphicx}
\usepackage{soul}
\usepackage[normalem]{ulem}
\usepackage{color}

\definecolor{notes}{HTML}{5C9384}

\newcommand{\clustername}{SDSS\,J1110$+$6459}
\newcommand{\clustershort}{SDSS\,J1110}
\newcommand{\galaxyname}{SGAS\,J111020.0$+$645950.8}
\newcommand{\galaxyshort}{SGAS\,J1110}
\newcommand{\lenstool}{\texttt{Lenstool}}
\newcommand{\wslap}{\texttt{WSLAP+}}

\newcommand{\clustz}{$z=0.659$}
\newcommand{\galz}{$z=2.481$}

\newcommand{\magtot}{$\mu_\mathrm{tot}=24.2^{+3.4}_{-1.2}$}

\newcommand{\masskpc}{$M(<250~\mathrm{kpc}) = 1.21^{+0.09}_{-0.04} \times 10^{14}~M_\odot$}
\newcommand{\masscrit}{$M(<A\mathrm{crit}) = 3.44^{+0.21}_{-0.25} \times 10^{13}~M_\odot$}
\newcommand{\areacrit}{$A_\mathrm{crit} = 367^{+21}_{-25}~\mathrm{arcsec}^2$}
\newcommand{\einstrad}{$\theta_\mathrm{E} = 10.8^{+0.3}_{-0.4}~\mathrm{arcsec}$}

\begin{document}

\title{LEGGOS II: A Strong Lens Model and Source-Plane Projection of the Clumpy Star-Forming Galaxy \galaxyname\ at \galz}


\author[0009-0004-9243-3459]{Pedram Abedi}
\affiliation{Department of Astronomy, University of Michigan, 1085 S. University Ave, Ann Arbor, MI 48109, USA}
\correspondingauthor{Pedram Abedi}
\email{pabedi@umich.edu}

\author[0000-0002-7559-0864]{Keren Sharon}
\affiliation{Department of Astronomy, University of Michigan, 1085 S. University Ave, Ann Arbor, MI 48109, USA}

\author[0000-0001-6251-4988]{Taylor A. Hutchison}
\altaffiliation{NASA Postdoctoral Fellow}
\affiliation{Astrophysics Science Division, Code 660, NASA Goddard Space Flight Center, 8800 Greenbelt Rd., Greenbelt, MD 20771, USA}
\affiliation{Department of Astronomy, University of Maryland, Baltimore Country, MD 21250, USA}
\affiliation{Center for Research and Exploration in Space Science and Technology, NASA/GSFC, Greenbelt, MD 20771 USA}

\author[0000-0003-1074-4807]{Matthew B. Bayliss}
\affiliation{Department of Physics, University of Cincinnati, Cincinnati, OH 45221, USA}

\author[0000-0001-5097-6755]{Michael Florian}
\affiliation{Steward Observatory, University of Arizona, 933 North Cherry Avenue, Tucson, AZ 85721, USA}
\affiliation{Eureka Scientific, 2452 Delmer Street Suite 100, Oakland, CA 94602-3017, USA}

\author[0000-0002-3475-7648]{Gourav Khullar}
\altaffiliation{Baum Postdoctoral Fellow for Innovative Astronomy}
\affiliation{Department of Astronomy, University of Washington, Physics-Astronomy Building, Box 351580, Seattle, WA 98195-1700, USA}
\affiliation{eScience Institute, University of Washington, Physics-Astronomy Building, Box 351580, Seattle, WA 98195-1700, USA}
\affiliation{Department of Physics and Astronomy and PITT PACC, University of Pittsburgh, Pittsburgh, PA 15260, USA}

\author[0000-0003-1370-5010]{Michael D. Gladders}
\affiliation{Department of Astronomy and Astrophysics, University of Chicago, 5640 South Ellis Avenue, Chicago, IL 60637, USA}
\affiliation{Kavli Institute for Cosmological Physics, University of Chicago, 5640 South Ellis Avenue, Chicago, IL 60637, USA}

\author[0009-0000-5333-9970]{Dylan Berry}
\affiliation{Department of Astronomy, University of Washington, Physics-Astronomy Building, Box 351580, Seattle, WA 98195-1700, USA}

\author[0000-0002-8261-9098]{Catherine Cerny}
\affiliation{Department of Astronomy, University of Michigan, 1085 S. University Ave, Ann Arbor, MI 48109, USA}

\author[0000-0003-2200-5606]{H{\aa}kon Dahle}
\affiliation{Institute of Theoretical Astrophysics, University of Oslo, P.O. Box 1029, Blindern, NO-0315 Oslo, Norway}

\author[0009-0009-4672-7807]{Aleena Ebey}
\affiliation{Department of Astronomy, University of Washington, Physics-Astronomy Building, Box 351580, Seattle, WA 98195-1700, USA}

\author[0000-0002-1728-8042]{Juliana S.M. Karp}
\affiliation{Department of Astronomy, University of Washington, Physics-Astronomy Building, Box 351580, Seattle, WA 98195-1700, USA}

\author[0000-0001-6505-0293]{Keunho Kim}
\affiliation{IPAC, California Institute of Technology, 1200 E. California Blvd., Pasadena CA, 91125, USA}

\author[0009-0000-3563-1695]{James W. Kulp}
\affiliation{Steward Observatory, University of Arizona, 933 North Cherry Ave., Tucson, AZ 85721, USA}

\author[0000-0003-3266-2001]{Guillaume Mahler}
\affiliation{STAR Institute, Quartier Agora - All\'ee du six Ao\^ut, 19c B-4000 Li\`ege, Belgium}

\author[0000-0002-7627-6551]{Jane R. Rigby}
\affiliation{Astrophysics Science Division, Code 660, NASA Goddard Space Flight Center, 8800 Greenbelt Rd., Greenbelt, MD 20771, USA}

\author[0000-0002-9204-3256]{T.\ Emil Rivera-Thorsen}
\affiliation{The Oskar Klein Centre, Department of Astronomy, Stockholm University, AlbaNova, 10691 Stockholm, Sweden}

\author[0009-0003-7031-2907]{Amritaansh Srivastava}
\affiliation{Steward Observatory, University of Arizona, 933 North Cherry Ave., Tucson, AZ 85721, USA}

\author[0000-0003-1815-0114]{Brian Welch}
\affiliation{International Space Science Institute, Hallerstrasse 6, 3012 Bern, Switzerland}

\author[0009-0005-8103-5823]{Alex Ross}
\affiliation{Department of Astronomy, University of Washington, Physics-Astronomy Building, Box 351580, Seattle, WA 98195-1700, USA}

\author[0000-0002-5293-3975]{Julissa Sarmiento}
\affiliation{Department of Physics and Astronomy, University of Pittsburgh, Pittsburgh, PA 15260, USA}

\begin{abstract}
Strong gravitational lensing by galaxy clusters combined with the resolution of JWST enables studies of star formation on $\sim10-100$ pc scales in galaxies at $z\sim2-4$. As part of the \textit{LEnsing and Galaxy Growth: Observing Substructures survey} (LEGGOS), we present an updated strong lensing model of the galaxy cluster \clustername\ (\clustz), which lenses the clumpy star-forming galaxy \galaxyname\ at \galz\ into a highly magnified giant arc. Using JWST NIRCam imaging, NIRSpec spectroscopy, and archival HST data, we confirm and refine the identification of three multiply imaged background sources, identify one new system, and map over 20 luminous regions between each image of the primary arc. Spectroscopy confirms that several previously ambiguous ``clumps" belong to the main arc at \galz. Despite the limited number of strongly lensed sources in the field, the resulting lens model has high precision, owing to the dense sampling of JWST-resolved clump constraints that tightly probe the lensing potential near the giant arc. The model yields a projected lens mass of \masskpc, an Einstein radius of \einstrad, and a total effective areal magnification of \magtot\ for the giant arc.  Individual clump magnifications span $\mu_\mathrm{clump}\sim4-19$, with fractional magnification uncertainties of $\sigma_\mu/|\mu_{\rm best}|\sim0.03-0.09$. We report a $\sim2.6-8.3\times$ improvement in magnification precision over previous models and estimate a $\sim7\%$ systematic uncertainty. This model enables ongoing and future LEGGOS studies of star formation, galaxy growth, and feedback within this highly magnified galaxy.
\end{abstract}


\section{Introduction} \label{sec:intro}

Strong gravitational lensing by massive galaxy clusters serves as a powerful tool for probing the internal structure of distant galaxies. The cluster potential distorts background sources into extended arcs and amplifies their observed flux. Such magnification allows observations to reach physical scales far below the native angular resolution of space-based imaging and spectroscopy, enabling detailed studies of galaxies during the epoch of peak cosmic star-formation, $z\sim1-3$ \citep[e.g.,][]{Livermore2012,Johnson2017,Rigby2017,Johnson_2021,Sharon2022,Welch2022,Vanzella2023,Rivera_Thorsen_2024,Hutchison2025}. 

The \textit{JWST LEGGOS survey} \citep[LEnsing and Galaxy Growth: Observing Substructures; GO-03843, PI: Bayliss; GO-04125, PIs: Florian,
Khullar;][]{khullar2026} combines the magnification from strong lensing with the unprecedented sensitivity of JWST in order to resolve clumpy star formation and structural evolution in highly magnified galaxies at $z\sim2-4$. Utilizing JWST NIRCam imaging, NIRSpec spectroscopy, archival HST data, and the magnification boost from strong lensing, LEGGOS achieves effective source-plane resolutions of $\sim10-100$ pc along the most highly magnified direction, with coarser resolution along the less-magnified direction. These observations enable the construction of a uniform framework for interpreting spatially resolved stellar populations across multiple highly magnified galaxies.

Central to LEGGOS is the development of accurate lens models for each system in the survey. Magnification directly affects inferred physical sizes, luminosities, and star-formation rates. Deflection fields determine the mapping between image-plane and source-plane morphology. Uncertainties in the lensing mass model propagate into derived physical quantities and into the interpretation of spatially resolved measurements. Homogeneous and well-constrained lensing solutions are therefore essential to enable robust comparisons across the LEGGOS sample.

\clustername, at \mbox{\clustz} (hereafter \clustershort;\autoref{fig:cstr}) is one of the clusters in the LEGGOS sample, lensing the background galaxy \galaxyname\ at \mbox{\galz} (hereafter \galaxyshort) into three images, forming a highly magnified giant arc roughly $\sim15''$ from the brightest cluster galaxy (BCG). This system was discovered as part of the Sloan Giant Arcs Survey \citep[SGAS,][]{sharon2020}, a systematic search for strongly lensed galaxies in the Sloan Digital Sky Survey (SDSS).  Spectroscopic follow-up in \citet{Johnson2017} confirmed a redshift of $z=2.4812\pm0.0005$ from nebular C\textsc{ii}] and C\textsc{iii}] emission lines, establishing \galaxyshort\ as a vigorously star-forming galaxy at cosmic noon. Lens modeling analysis by \citet{Johnson2017} further demonstrated the power of this system to resolve star-forming structure at $\lesssim 30$–$50~\mathrm{pc}$ scales using HST imaging \citep{Johnson2017_ii}.



JWST's enhanced spatial resolution and wavelength coverage enables detailed mapping of clumpy star formation across the arc \citep[e.g.,][]{Mahler2023}. With the deep JWST NIRCam imaging and NIRSpec spectroscopy, \galaxyshort\ can now be studied with substantially improved sensitivity compared to previous HST-only analyses. In this paper, we fully leverage these data and construct an updated strong lensing solution, taking advantage of the superior JWST imaging and spectroscopy. The new model provides improved magnification and deflection maps calibrated for JWST-based analyses. The model was constructed using similar methods to other LEGGOS targets \citep[][]{khullar2026,Welch2026,sharon2022_1226,Sharon2022,sharon2020}, ensuring consistency across the survey. The goal is to provide the lensing infrastructure necessary for subsequent LEGGOS science analyses: a well-constrained cluster mass model, calibrated magnification and deflection maps, and source-plane projections of the primary lensed galaxy. We focus exclusively on the construction and validation of the lensing solution and do not interpret the internal star-forming structure of \galaxyshort, which will be presented in forthcoming papers.

In \autoref{sec:data}, we describe the JWST and archival HST observations and data reduction used in this work. In \autoref{sec:lensmodel}, we present the strong lens modeling methodology, including the identification of multiply imaged systems and computation of the mass model and magnifications. The results of the lens model, including the mass distribution, magnification measurements, and source-plane projections, are presented in \autoref{sec:results}. We discuss the implications of these results, assess systematic uncertainties, and compare to previous work in \autoref{sec:discussion}, and summarize our conclusions in \autoref{sec:conclusion}. We provide model predictions of time delays and photometry of secondary arcs in \autoref{sec:appendix} and \autoref{sec:appendix_magnitdues}.

Throughout this paper, we assume a flat cosmology with $\Omega_M=0.3$, $\Omega_\Lambda=0.7$, and $H_0=70~\text{km}~\text{s}^{-1}~\text{Mpc}^{-1}$. An angular size of $1''$ corresponds to a physical distance of $6.97~\mathrm{kpc}$ at the cluster redshift $z = 0.659$ and $8.085~\mathrm{kpc}$ at the redshift of the giant arc $z = 2.481$. All magnitudes are reported in the AB system.

\section{Observations and Data Reduction} \label{sec:data}
JWST observations of \clustershort\ were obtained as part of the LEGGOS survey \citep{khullar2026}. In this work, we utilize the NIRCam imaging products for lens modeling and NIRSpec spectroscopy to confirm redshifts of several clumps in the the primary arc. Archival HST imaging and previous results from ground based data \citep{Johnson2017,sharon2020} are included to aid in the identification of image constraints, spectroscopic source redshifts, and spectroscopic cluster member redshifts.

\subsection{JWST NIRCam Imaging}
\clustershort\ was observed with the JWST Near Infrared Camera (NIRCam) as part of the LEGGOS survey (GO-03843, PI: Bayliss; GO-04125, PIs: Florian, Khullar) in 10 filters: F070W, F090W, F115W, F277W, F356W, and F444W (GO-03843), and F150W, F182M, F200W, and F480M (GO-04125). \autoref{fig:cstr} shows color images of the cluster core and zoom-in on lensing features from a subset of these filters. Together, these filters span rest-frame $0.13–1.3\mathrm{\mu m}$ (UV through near-IR) at \galz, providing a broad spectral baseline for constraining galaxy morphology and stellar populations. The filter strategy was designed to sample stellar continuum while avoiding strong nebular emission features where possible \citep{khullar2026}.

All imaging was obtained using a four-point sub-pixel dither pattern to improve PSF sampling and mitigate detector artifacts. NIRCam observes in simultaneous short-wavelength (SW) and long-wavelength (LW) channel pairs; integration times for the GO-03843 observations were set by the LW bands to achieve signal-to-noise ratios $\gtrsim 10$ for the faintest resolved clumps identified in pre-existing HST imaging. The total integration times for these observations were 2490s in F070W/F444W, 1460s in F090W/F356W, and 1245s in F115W/F277W. Additional NIRCam observations from GO-04125 provide imaging in F150W, F182M, F200W, and F480M. The total integration times of the NIRCam observations from GO-04125 were 858.9s in F150W, F182M, and F200W, and 2576.8s in F480M. Based on the JWST Exposure Time Calculator (ETC), the observations have approximate $5\sigma$ point-source depths of 27.5, 27.3, 27.3, 27.3, 26.9, 27.4, 27.5, 27.9, 27.4, and 26.4 mag in F070W, F090W, F115W, F150W, F182M, F200W, F277W, F356W, F444W, and F480M, respectively.

The calibrated exposures were processed using the JWST calibration pipeline (v1.15.1) with CRDS context version 1298 and combined into final science-ready mosaics. During reduction, the images were corrected for detector artifacts, background structure, and residual instrumental signatures. In addition, a dedicated correction was applied to mitigate detector $1/f$ noise patterns---low-frequency correlated noise introduced by the detector readout electronics, which are not fully removed by the default JWST pipeline and can introduce large-scale background variations across the detector \citep{Rauscher24_NSClean}. The custom correction suppresses correlated striping patterns and improves the reliability of faint morphological structures used as constraints in the lens model. The final mosaics were drizzled onto a common $0\farcs03$ pixel scale. Further details of the observing strategy and reduction procedures are presented in \citet{khullar2026}. 

\subsection{JWST NIRSpec Spectroscopy}
Spectroscopic observations are used in this work to spectroscopically confirm a few clumps in \galaxyshort, adding valuable constraints to the lensing analysis. The central image of \galaxyshort\ (labeled 1.1 in \autoref{fig:cstr}) was observed with the JWST Near Infrared Spectrograph (NIRSpec) using integral field spectroscopy (IFS) as part of GO-03843. The IFS pointing was centered at $\mathrm{R.A.}=167.5831^\circ,\,\mathrm{Decl.}=+64.9978^\circ$, covering most of image 1.1 \citep[for the observed NIRSpec footprint, see][]{khullar2026}. A continuum map constructed from the NIRSpec IFS observations is shown in the right panel of \autoref{fig:s2_specz}. The observations were obtained in two high-resolution grating configurations: G140H/F100LP and G235H/F170LP. At the redshift of the background source (\galz), these settings provide continuous spectral coverage over rest-frame wavelengths of approximately $0.3–0.9 \mu\mathrm{m}$.

Exposure times were determined using the JWST ETC to ensure detection of key nebular emission lines from the lensed galaxy. The total integration times were 10737s in G140H/F100LP and 7119s in G235H/F170LP. The observations were divided into 16 and 8 exposures, respectively, and obtained using an 8-point \texttt{CYCLING-SMALL} dither pattern. This combined observing strategy mitigates the effects of cosmic rays, limits individual exposure times, and improves spatial sampling, enabling the construction of final spectral data cubes with effective spaxel sizes smaller than the native $0\farcs1$ NIRSpec IFS sampling.

The data were processed using the JWST calibration pipeline (v1.20.2) with CRDS context version 1466, together with the custom post-processing described in \citet{Hutchison24baryonsweep}, to produce science-ready spectral cubes. For further details of the observing strategy, data reduction, and spectroscopic analysis see \citet{khullar2026} and \citet{Welch2026}.

\subsection{Archival HST Imaging}
We incorporate archival HST imaging of \clustershort\ obtained with WFC3 as part of HST-GO-13003 (PI: Gladders) in four broadband filters: F390W and F606W (UVIS) and F105W and F160W (IR). This combination provides rest-frame ultraviolet through optical coverage of the lensed galaxy at \galz\ \citep{Johnson2017}. The archival data were reduced using the standard WFC3 calibration pipeline and combined with \texttt{AstroDrizzle} onto a common $0\farcs03$ pixel scale. Full details of the reduction procedure are described in \citet{Johnson2017} and the reduced data were made publicly available by \cite{sharon2020}. The final mosaics reach $5\sigma$ limiting magnitudes of $m=26.43$, $26.47$, $25.36$, and $25.68$ in F390W, F606W, F105W, and F160W, respectively. This work primarily uses the UVIS filters, since the new NIRCam imaging supersedes the WFC3-IR both in depth and resolution.

\section{Strong Lensing Methods} \label{sec:lensmodel}
\subsection{Previous Lensing Work}\label{subsec:prev_work}
\cite{Johnson2017} presented a lensing analysis of \clustershort, based on  HST imaging and ground-based spectroscopy. Following the methodology of \cite{Jullo_Kneib}, they employed a hybrid parametric/freeform modeling approach within \lenstool\ \citep{Jullo2007} that combines standard parametric cluster and galaxy halos with a flexible non-parametric correction. They modeled all mass components as truncated pseudo-isothermal elliptical mass distributions, including both the primary cluster-scale halos and cluster-member galaxies, and a multiscale grid of localized mass perturbations. Eight realizations of hybrid models were presented in order to reflect different modeling choices.

The models used strong lensing constraints identified in the HST imaging. For the primary arc (source A in \citealt{Johnson2017}), six clumps were identified in each of the three merging images and used as lensing constraints. Additional multiply imaged systems included five images of a second background source (source B) and two images of a third source (source C). They treated three clumps projected near the primary arc as possible separate sources (sources D, E, and F). The redshifts of all systems except source A were not spectroscopically confirmed and therefore allowed to vary as free parameters in the model. 

Using the parametric/freeform hybrid framework, \citet{Johnson2017} calculated the total areal magnification of the entire arc from each of their eight best-fit models. They reported $\mu_\mathrm{tot}=28\pm8$, which represents the median and median absolute deviation of the eight best-fit magnification values. Among their eight models, Model 2 (hereafter J17-M2) yielded the lowest image-plane RMS, $\mathrm{RMS_{i}}=0\farcs10$, and was adopted for all subsequent source-plane analyses \citep{Johnson2017,Rigby2017,Johnson2017_ii}. \citet{Johnson2017} reconstructed the source-plane structure of the most magnified image of the primary arc with forward modeling, and subsequent analysis in \citet{Johnson2017_ii} measured sizes and star formation rates of star-forming clumps down to $\lesssim30-50~\mathrm{pc}$.

\subsection{Identification of Multiply Imaged Systems}
Our search for multiple images builds upon the systems previously identified by \cite{Johnson2017} from HST imaging based on morphological similarity and consistent colors across WFC3 filters. The deeper and higher-resolution JWST/NIRCam imaging, along with its expanded wavelength coverage, allows us to confirm these earlier identifications and refine the morphology of several systems. We identify new substructures within the images that provide additional constraints for the lens model, and identify a new lensed source in this field. 

In total, we identify four background sources producing 12 secure images. Three of these sources contain visible substructure that is used as additional constraints when seen in more than one of their respective images. 

\begin{figure*}
\centering
    \includegraphics[width=1\linewidth]{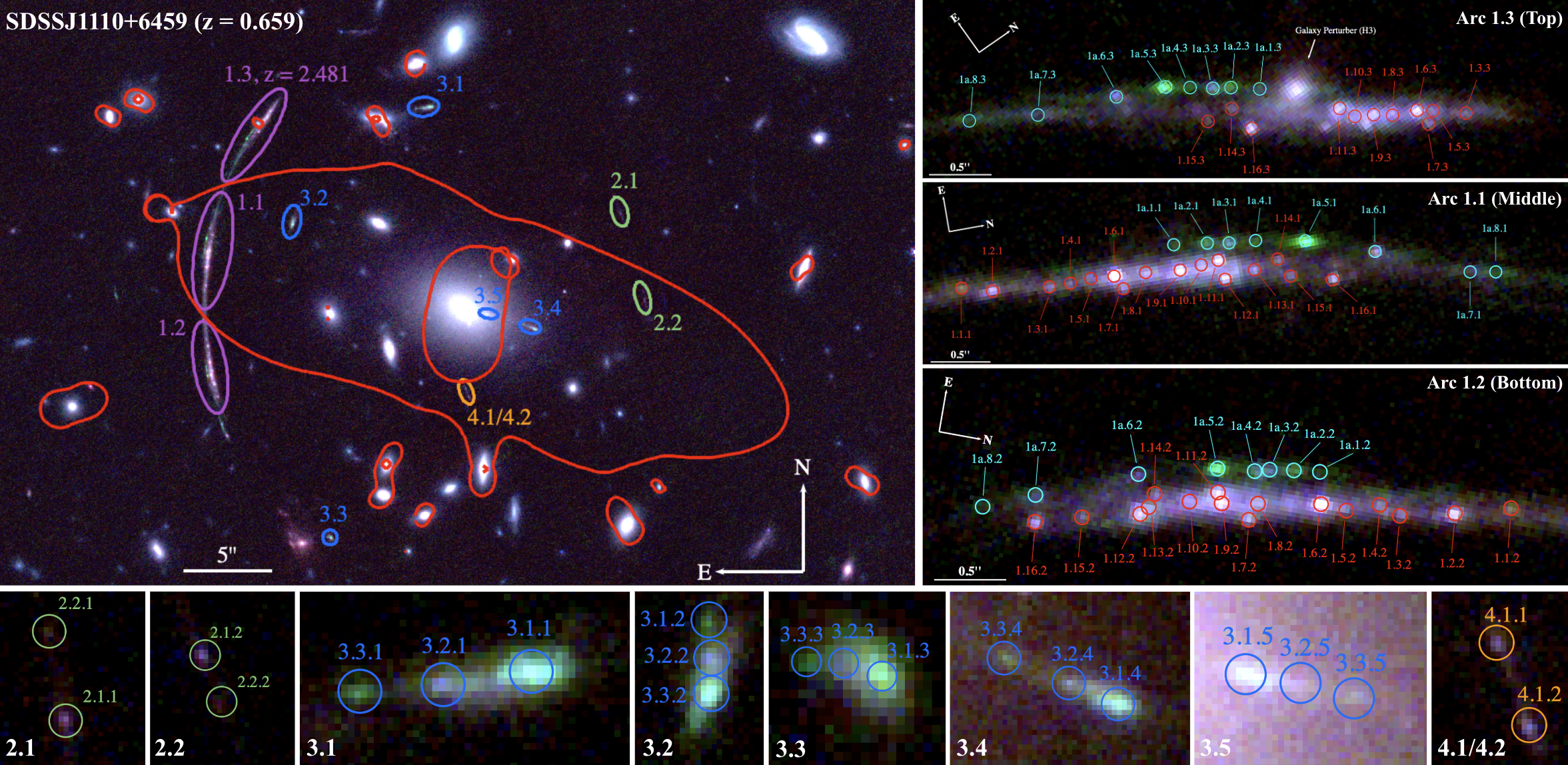}
    \caption{(Top left) JWST NIRCam color image of \clustershort\ constructed from F480M  (red), F182M (green), and F150W (blue). The critical curve for a source at \galz\ is shown in red. Multiply imaged systems used in the lens model are labeled. (Top right) Close-up view of the three images of the primary arc, shown using a color composite of F200W (red), F182M (green), and F115W (blue). The central image (1.1) exhibits an expected parity flip relative to the other two images. (Bottom) Cut-out images of additional multiply imaged systems used as constraints, displayed using the same color combination as the top right. Each cut-out shows the corresponding image positions and identified clumps used in the model. The orientation of these images match that of the top left image.} 
    \label{fig:cstr}
\end{figure*}

The positions of constraints are centered on distinct morphological features, which are identified and matched by eye, as the magnifications and distortions of these features can vary between images. These positions are tabulated in \autoref{tab:constraints}. Image IDs follow the convention \textit{family ID, clump ID, multiple image ID}. All constraints are shown in \autoref{fig:cstr}. The systems are described in the following subsections. 

\begin{longdeluxetable}{cccc}
\tablecaption{Lensing Constraints in \clustername \label{tab:constraints}}
\tablehead{
\colhead{Arc ID} &
\colhead{R.A.} &
\colhead{Decl.} &
\colhead{$\mu_{\text{clump}}$} \\[-6pt]
\colhead{} &
\colhead{(J2000, deg)} &
\colhead{(J2000, deg)} &
\colhead{}
}
\startdata
1.1.1 & 167.5833828 & 64.9969354 & $15.8^{+3.9}_{-1.5}$ \\
1.1.2 & 167.5833299 & 64.9961624 & $17.6^{+3.8}_{-1.8}$ \\
1.2.1 & 167.5833367 & 64.9970053 & $13.0^{+3.0}_{-1.2}$ \\
1.2.2 & 167.5832618 & 64.9960581 & $13.8^{+2.9}_{-1.4}$ \\
1.3.1 & 167.5833058 & 64.9971361 & $10.61^{+2.41}_{-0.99}$ \\
1.3.2 & 167.5832071 & 64.9959544 & $11.2^{+2.2}_{-1.1}$ \\
1.3.3 & 167.5808037 & 64.9998098 & $3.79^{+0.30}_{-0.12}$ \\
1.4.1 & 167.5833054 & 64.9971857 & $10.09^{+2.27}_{-0.93}$ \\
1.4.2 & 167.5832397 & 64.9959125 & $10.12^{+1.93}_{-0.93}$ \\
1.5.1 & 167.5833100 & 64.9972343 & $9.71^{+2.16}_{-0.89}$ \\
1.5.2 & 167.5831876 & 64.9958507 & $9.21^{+1.71}_{-0.83}$ \\
1.5.3 & 167.5809125 & 64.9997533 & $4.02^{+0.33}_{-0.12}$ \\
1.6.1 & 167.5833011 & 64.9972884 & $9.22^{+2.00}_{-0.82}$ \\
1.6.2 & 167.5831967 & 64.9958010 & $8.42^{+1.51}_{-0.74}$ \\
1.6.3 & 167.5809621 & 64.9997251 & $4.16^{+0.35}_{-0.13}$ \\
1.7.1 & 167.5832256 & 64.9973037 & $8.48^{+1.72}_{-0.69}$ \\
1.7.2 & 167.5830688 & 64.9956694 & $7.12^{+1.17}_{-0.58}$ \\
1.7.3 & 167.5808696 & 64.9997283 & $4.12^{+0.34}_{-0.13}$ \\
1.8.1 & 167.5832909 & 64.9973611 & $8.69^{+1.79}_{-0.74}$ \\
1.8.2 & 167.5831478 & 64.9956815 & $7.10^{+1.18}_{-0.58}$ \\
1.8.3 & 167.5810142 & 64.9996733 & $4.50^{+0.39}_{-0.16}$ \\
1.9.1 & 167.5832717 & 64.9974403 & $8.16^{+1.58}_{-0.65}$ \\
1.9.2 & 167.5831209 & 64.9956122 & $6.49^{+1.01}_{-0.51}$ \\
1.9.3 & 167.5810747 & 64.9996399 & $4.86^{+0.44}_{-0.21}$ \\
1.10.1 & 167.5832804 & 64.9974903 & $8.02^{+1.48}_{-0.61}$ \\
1.10.2 & 167.5831037 & 64.9955505 & $6.01^{+0.89}_{-0.46}$ \\
1.10.3 & 167.5811242 & 64.9996039 & $5.57^{+0.58}_{-0.31}$ \\
1.11.1 & 167.5832892 & 64.9975297 & $7.92^{+1.40}_{-0.57}$ \\
1.11.2 & 167.5831675 & 64.9956014 & $6.33^{+0.98}_{-0.50}$ \\
1.11.3 & 167.5812033 & 64.9995864 & $6.48^{+0.87}_{-0.49}$ \\
1.12.1 & 167.5831800 & 64.9975381 & $7.24^{+1.25}_{-0.47}$ \\
1.12.2 & 167.5830088 & 64.9954611 & $5.49^{+0.76}_{-0.39}$ \\
1.13.1 & 167.5832032 & 64.9976089 & $7.16^{+1.18}_{-0.42}$ \\
1.13.2 & 167.5830472 & 64.9954750 & $5.54^{+0.77}_{-0.40}$ \\
1.14.1 & 167.5832396 & 64.9976672 & $7.18^{+1.16}_{-0.35}$ \\
1.14.2 & 167.5831108 & 64.9954819 & $5.53^{+0.77}_{-0.40}$ \\
1.14.3 & 167.5815292 & 64.9993900 & $6.58^{+0.79}_{-0.30}$ \\
1.15.1 & 167.5831338 & 64.9976898 & $6.70^{+1.02}_{-0.31}$ \\
1.15.2 & 167.5829476 & 64.9953518 & $4.91^{+0.62}_{-0.32}$ \\
1.15.3 & 167.5815454 & 64.9993311 & $6.62^{+0.75}_{-0.26}$ \\
1.16.1 & 167.5830760 & 64.9977836 & $6.45^{+1.02}_{-0.22}$ \\
1.16.2 & 167.5828899 & 64.9952642 & $4.52^{+0.54}_{-0.28}$ \\
1.16.3 & 167.5813836 & 64.9994011 & $10.2^{+2.5}_{-1.2}$ \\
\hline
1a.1.1 & 167.5834145 & 64.9974367 & $9.40^{+1.98}_{-0.87}$ \\
1a.1.2 & 167.5833410 & 64.9957875 & $7.86^{+1.38}_{-0.67}$ \\
1a.1.3 & 167.5815290 & 64.9994656 & $7.45^{+1.19}_{-0.56}$ \\
1a.2.1 & 167.5833912 & 64.9975137 & $8.77^{+1.69}_{-0.73}$ \\
1a.2.2 & 167.5833267 & 64.9957382 & $7.35^{+1.25}_{-0.62}$ \\
1a.2.3 & 167.5816220 & 64.9994154 & $6.14^{+0.67}_{-0.25}$ \\
1a.3.1 & 167.5833721 & 64.9975627 & $8.37^{+1.49}_{-0.62}$ \\
1a.3.2 & 167.5833104 & 64.9956920 & $6.92^{+1.15}_{-0.57}$ \\
1a.3.3 & 167.5816732 & 64.9993818 & $6.01^{+0.62}_{-0.22}$ \\
1a.4.1 & 167.5833600 & 64.9976234 & $8.00^{+1.38}_{-0.52}$ \\
1a.4.2 & 167.5832921 & 64.9956637 & $6.69^{+1.09}_{-0.54}$ \\
1a.4.3 & 167.5817442 & 64.9993418 & $6.03^{+0.63}_{-0.20}$ \\
1a.5.1 & 167.5833121 & 64.9977380 & $7.26^{+1.25}_{-0.28}$ \\
1a.5.2 & 167.5832734 & 64.9955928 & $6.14^{+0.95}_{-0.47}$ \\
1a.5.3 & 167.5818194 & 64.9992976 & $6.19^{+0.66}_{-0.20}$ \\
1a.6.1 & 167.5831876 & 64.9978918 & $6.74^{+1.20}_{-0.19}$ \\
1a.6.2 & 167.5831862 & 64.9954445 & $5.25^{+0.70}_{-0.37}$ \\
1a.6.3 & 167.5819251 & 64.9991955 & $6.91^{+0.80}_{-0.25}$ \\
1a.7.1 & 167.5829871 & 64.9981002 & $8.31^{+1.24}_{-0.20}$ \\
1a.7.2 & 167.5830104 & 64.9952560 & $4.43^{+0.52}_{-0.27}$ \\
1a.7.3 & 167.5820823 & 64.9990304 & $9.07^{+1.22}_{-0.31}$ \\
1a.8.1 & 167.5829637 & 64.9981586 & $9.50^{+1.26}_{-0.24}$ \\
1a.8.2 & 167.5829170 & 64.9951600 & $4.10^{+0.45}_{-0.23}$ \\
1a.8.3 & 167.5822633 & 64.9988986 & $12.32^{+1.92}_{-0.34}$ \\
\hline
2.1.1 & 167.5671317 & 64.9967681 & $4.75^{+1.86}_{-0.53}$ \\
2.1.2 & 167.5680233 & 64.9981308 & $6.32^{+1.13}_{-0.74}$ \\
2.2.1 & 167.5671825 & 64.9969150 & $6.28^{+2.81}_{-0.84}$ \\
2.2.2 & 167.5679562 & 64.9980419 & $7.14^{+1.48}_{-0.95}$ \\
\hline
3.1.1 & 167.5750392 & 64.9997917 & $4.45^{+0.85}_{-0.62}$ \\
3.1.2 & 167.5801301 & 64.9979502 & $2.98^{+0.48}_{-0.14}$ \\
3.1.3 & 167.5786870 & 64.9930606 & $2.40^{+0.40}_{-0.15}$ \\
3.1.4 & 167.5711650 & 64.9963143 & $2.17^{+0.46}_{-0.29}$ \\
3.1.5 & 167.5729921 & 64.9965587 & $0.54^{+0.15}_{-0.07}$ \\
3.2.1 & 167.5753218 & 64.9997737 & $4.72^{+0.94}_{-0.69}$ \\
3.2.2 & 167.5801276 & 64.9980058 & $3.15^{+0.49}_{-0.14}$ \\
3.2.3 & 167.5787746 & 64.9930729 & $2.39^{+0.39}_{-0.15}$ \\
3.2.4 & 167.5713475 & 64.9963533 & $2.43^{+0.48}_{-0.36}$ \\
3.2.5 & 167.5728342 & 64.9965431 & $0.83^{+0.21}_{-0.11}$ \\
3.3.1 & 167.5755820 & 64.9997597 & $4.86^{+0.97}_{-0.68}$ \\
3.3.2 & 167.5801373 & 64.9980607 & $3.35^{+0.51}_{-0.15}$ \\
3.3.3 & 167.5788458 & 64.9930728 & $2.37^{+0.37}_{-0.14}$ \\
3.3.4 & 167.5716031 & 64.9963892 & $3.07^{+0.65}_{-0.55}$ \\
3.3.5 & 167.5726307 & 64.9965252 & $1.54^{+0.50}_{-0.19}$ \\
\hline
4.1.1 & 167.5737458 & 64.9954003 & $24^{+18}_{-12}$ \\
4.1.2 & 167.5736254 & 64.9952581 & $24.4^{+...}_{-4.6}$ \\
\enddata
\tablenotetext{}{
\parbox{0.90\columnwidth}{
The redshift of source 1 and 1a are fixed to the spectroscopic redshift of the primary arc, \galz\ (see \autoref{sec:primary_arc}).
}}
\vspace{4pt}
\tablenotetext{}{
\parbox{0.90\columnwidth}{
Clumps 1a.6, 1a.5, and 1a.3 correspond to D, E, and F in \citet{Johnson2017}.
}}
\vspace{4pt}
\tablenotetext{}{
\parbox{0.90\columnwidth}{
Systems 21-22 and 31-33 correspond to sources C and B in \citet{Johnson2017}.
}}
\vspace{4pt}
\tablenotetext{}{
\parbox{0.90\columnwidth}{
Model-predicted redshifts: $z_{\mathrm{model},2}=2.96^{+0.46}_{-0.28}$, $z_{\mathrm{model},3}=3.07^{+0.22}_{-0.13}$, $z_{\mathrm{model},4}=4.80^{+0.16}_{-1.57}$.
}}
\vspace{4pt}
\tablenotetext{}{
\parbox{0.90\columnwidth}{
Central $\mu_{\mathrm{clump}}$ values correspond to the best-fit model, with uncertainties derived from the 95\% confidence interval of 300 models randomly sampled from the posterior distribution.
}}
\vspace{4pt}
\tablenotetext{}{
\parbox{0.90\columnwidth}{
Arc IDs follow the convention: family ID, clump ID, multiple image ID.
}}
\vspace{4pt}
\tablenotetext{}{
\parbox{0.90\columnwidth}{
The upper limit on magnification for arc 4.1.2 is unconstrained.
}}
\end{longdeluxetable}

\subsubsection{The Primary Arc \galaxyname}
\label{sec:primary_arc}
\galaxyshort\ is the primary arc, stretching $\sim17''$ across and is $\sim15''$ from the BCG consisting of three, partially merging images (\autoref{fig:cstr}). \citet{Johnson2017} identified and mapped six clumps in each image (labeled Aa-Af in their paper). We identify 19 clumps that can be mapped between all three multiple images of the arc system, and an additional five clumps that can be matched between the middle and southern images, but are not identified in the less magnified northern image. Of these 24 clumps, eight are identified as a sub-system of source 1, labeled 1a (see below). The middle image exhibits the highest magnification and was used to identify clumps before searching for their counterparts in the other two images.

We note two bright clumps south of $1.2$ with colors similar to the primary arc. We concur with the interpretation of \cite{Johnson2017} that these may be an extension of the primary arc, but are not multiply imaged due to their location outside the lensing caustic in the source-plane.

\textit{System 1a: edge clumps---} \citet{Johnson2017} identified three clumps (D, E, and F) along the edge of the primary arc that appear spatially offset from the main arc structure and significantly different in color. Due to the limited spatial resolution of ground-based spectroscopy, their redshifts could not be independently constrained, and they were therefore modeled as a separate system in \citet{Johnson2017}. 
With the deeper JWST imaging, we identify eight knots associated with this structure that can be matched across the three images ($1\text{a.*}.1$, $1\text{a.*}.2$, and $1\text{a.*}.3$). In our labeling scheme, clumps D, E, and F correspond to 1a.6.*, 1a.5.*, and 1a.3.*, respectively.

\begin{figure*}
\centering
    \includegraphics[width=1\linewidth]{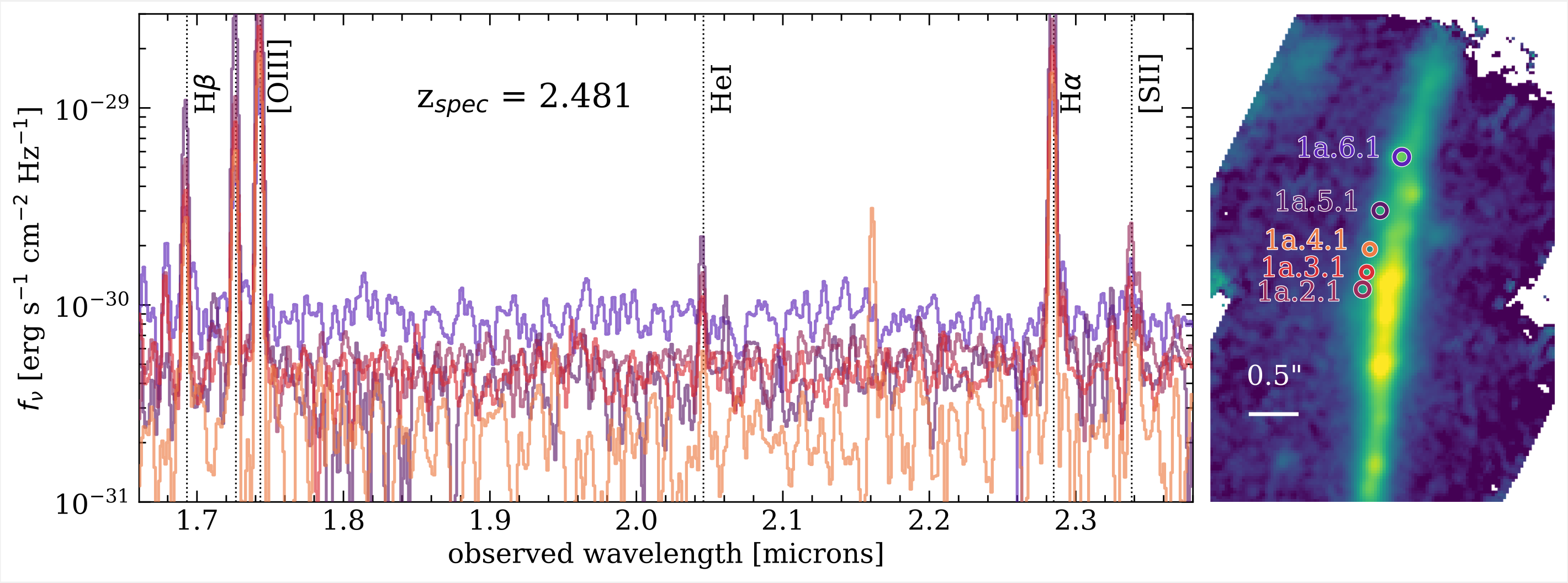}
    \caption{JWST/NIRSpec spectroscopy of representative clumps associated with the edge structure, system 1a.
    (Left) One-dimensional spectra of five clumps extracted from NIRSpec data. Prominent emission lines are detected at a consistent redshift of \galz, matching the primary arc.
    (Right) NIRSpec continuum map of the arc with the locations of the extracted clumps overlaid. The consistent redshift across these distinct regions confirms the edge structure arises from the same background galaxy as the primary arc.} 
    \label{fig:s2_specz}
\end{figure*}

Using JWST/NIRSpec observations, we confirm that these clumps lie at the same redshift as the primary arc, from prominent emission features detected in their spectra, including H$\beta$, [O\,III], He\,I, H$\alpha$, and [S\,II] (\autoref{fig:s2_specz}). The strong [O\,III] emission traces ionized gas associated with star formation and likely contributes significantly to the observed brightness of these clumps in the F182M bandpass, explaining their distinct color in the JWST filter combination shown in \autoref{fig:cstr}. A detailed analysis and interpretation of these spectroscopic observations will be presented in a forthcoming paper (Hutchison et al., 2026, in preparation).

For clarity in discussion, we retain a separate label for these edge clumps (system 1a), even though they are part of the same source as the primary arc. Their apparent spatial offset between multiple images arises from strong lensing shear along the source, which differentially stretches and displaces features depending on their position with respect to the inner or outer edge of the giant arc (where ``outer edge'' here means the edge of the arc farther from the BCG).

\subsubsection{Identification of Secondary Arcs}
\label{subsubsec:secondary_arc_ids}

In addition to the primary arc system, we use several secondary arc systems as modeling constraints. Secondary arcs are identified through visual inspection of the imaging data, guided by similarities in color and morphology among candidate counter images.

Arcs $2.1$ and $2.2$ correspond to system C in \citet{Johnson2017}. These tangential arcs are located $10\farcs1$ west and $10\farcs3$ northwest of the BCG, respectively. The lens model predicts a third counter image south of the BCG; however, due to its distance from the critical curve in this region, we expect it to be only weakly magnified, making it difficult to identify confidently. We tested several candidate counter images identified in the JWST imaging, but these resulted in poorer model reproductions of arcs $2.1$ and $2.2$. Candidate counter images proposed by \citet{Johnson2017} could not be confirmed. We therefore adopt only the two clearly identified images as constraints. The JWST imaging resolves two distinct clumps in each arc, which serve as positional constraints, improving upon the constraints presented by \citet{Johnson2017}.

System 3 corresponds to system B in \citet{Johnson2017}. This system consists of three tangential arcs ($3.1-3.3$) located $11\farcs5$ north, $10\farcs8$ northeast, and $14\farcs9$ from the BCG, and two radial arcs ($3.4$ and $3.5$) located $4\farcs1$ and $1\farcs2$ west of the BCG. We improve upon the previous model by robustly identifying three clumps in each of the five images, which are used as additional constraints in the lens model.

Arcs $4.1$ and $4.2$ correspond to a newly identified radial arc system, located $4\farcs4$ and $5\farcs0$ south of the BCG. These arcs are faint in the HST imaging but become pronounced in the JWST data. The lens model predicts a third less-magnified counter image north of the BCG; however, no convincing counterpart with consistent morphology or color could be identified in the available imaging depth.

\subsubsection{Photometry of Secondary Arcs}

To add confidence in the secondary arc identifications, we compare the colors of the multiple images using aperture photometry measured in the four HST and ten JWST/NIRCam filters. The imaging was PSF-matched to the F480M data before measuring the photometry, and foreground light was modeled and subtracted where necessary. The resulting magnitudes are presented in  \autoref{sec:appendix_magnitdues}. Because gravitational lensing is achromatic, multiple images of the same source are expected to exhibit similar colors, while their apparent brightness may differ because of their different magnifications. The images within each system generally show similar photometric trends across filters. Images projected near the BCG are particularly susceptible to systematic uncertainties from foreground-light subtraction and source deblending. Together, the photometry, image morphologies, and lens-model predictions solidify the secondary arc identifications.

We attempted to estimate photometric redshifts for each multiple image in systems 2, 3, and 4 using methods similar to those described in \citet{cerny2025slice}. We fit the aperture photometry measured across the available HST and JWST filters with stellar population synthesis models using the \texttt{Prospector} \citep{Johnson_2021} framework within a Bayesian inference scheme to estimate redshift probability distributions.

The resulting photometric redshift posteriors are broad, and although the arcs are robustly identified as multiple images, in some cases images of the same system do not converge to consistent photometric redshift solutions. Systems 2, 3, and 4 exhibit non-zero probability over wide redshift intervals of [1–10], [2.3–2.7], and [2.7–3.7], respectively, indicating that the photometric redshifts are poorly constrained. To be conservative and avoid bias from catastrophic failures in the fit \citep{Sharon2023}, we use the photometric redshift estimates only to inform broad, physically motivated priors on the source redshifts in the lens model optimization, rather than adopting them as direct constraints. The adopted priors are chosen to encompass the most plausible redshift range for strongly lensed sources given the lensing configuration, while excluding extreme tails in the photometric posteriors that are unlikely to be physical. Specifically, we allow the redshift of systems 2 and 3 to vary within a flat prior of $2 < z < 5$, and system 4 within $2.8 < z < 5$. These ranges bracket the primary photo-$z$ solutions while maintaining sufficient flexibility in the lens model and avoiding influence from poorly constrained photometric estimates.

\subsection{Identification of Cluster Member Galaxies}

Cluster member galaxies contribute localized, galaxy-scale perturbations to the overall cluster potential. We construct a catalog of cluster members using a red-sequence selection criterion \citep{GladdersYee2000} anchored by spectroscopically confirmed members from \cite{Johnson2017}. Source detection and photometry are performed on WFC3/IR F105W and WFC3/UVIS F606W using \textsc{Source Extractor} \citep{Bertin1996} in dual-image mode, with F105W adopted as the detection band. This filter combination brackets the 4000~\AA\ break at the cluster redshift, with F606W sampling the blue side and partially overlapping the break, and F105W sampling redward of the break, providing a well-defined color for red-sequence selection in F606W$-$F105W versus F105W color-magnitude space. We adopt HST-only photometry for this step to be consistent with previous work \citep{Johnson2017,Sharon2023}.

\begin{figure}
\centering
    \includegraphics[width=1\linewidth]{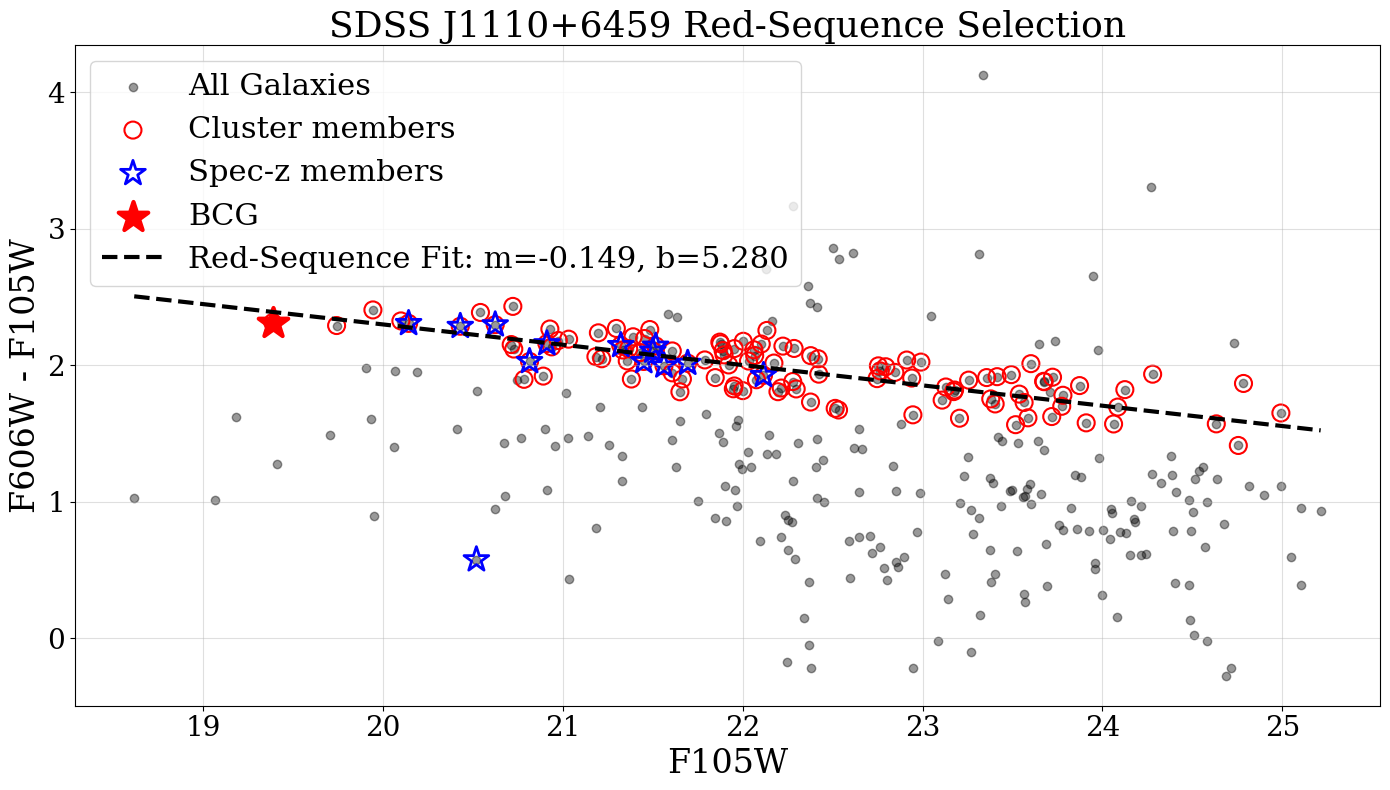}
    \caption{Red-sequence selection of SDSS J1110 cluster members. Blue stars are spectroscopically confirmed cluster members used to inform the red-sequence fit. The bluer confirmed member is suspected to be a jellyfish galaxy falling into the cluster. Cluster members (red circles) are selected out to $3.5\sigma$} 
    \label{fig:galcat}
\end{figure}

Spectroscopically-confirmed cluster members \citep[17 galaxies,][]{Johnson2017} highlight the red sequence in F606W$-$F105W versus F105W color-magnitude space, which can be seen in \autoref{fig:galcat}. To determine the red-sequence relation, we perform a bootstrap resampling of the spectroscopic members with replacement. The procedure is iterated until the median slope and intercept converge to within 1\% between successive iterations. Residuals are computed relative to the best-fit relation, and a single $3\sigma$ clipping is applied to remove outliers (star-forming cluster member galaxies, which are bluer than the red sequence). The bootstrap is then repeated using the clipped sample to derive the final red-sequence relation. We select cluster members as galaxies lying within $3.5\sigma$ of this relation, where $\sigma$ is the scatter in color of the clipped spectroscopically confirmed cluster member sample around the best-fit red-sequence line. This threshold is chosen to ensure inclusion of spectroscopically confirmed members that lie near the boundary of the red sequence, while avoiding the increased contamination introduced by broader selections. We perform a final visual inspection to remove contaminants such as arcs and substructure in foreground galaxies with color similar to the red sequence. The final red-sequence catalog contains 112 galaxies.

\subsection{Lens Modeling}\label{subsec:lensModeling}
We model the lensing potential of \clustershort\ using the publicly available software \lenstool\ \citep{Jullo2007}. The modeling framework assumes that the lens-plane can be represented by a combination of parametric halos; it employs a Bayesian Markov Chain Monte Carlo (MCMC) sampling technique to explore and optimize the parameter space. To maintain consistency with the broader LEGGOS survey lens modeling framework, we adopt the traditional parametric \lenstool\ modeling approach rather than the hybrid method used in the analysis of \cite{Johnson2017}. 

The lens modeling procedure is iterative. We begin with an initial set of constraints and a simple mass model based on initial guesses of the cluster mass distribution. This preliminary model is progressively refined as additional constraints from newly identified multiple images of clumps are incorporated and the mass distribution is adjusted. Model outputs predict potential new multiple images, which are searched for and incorporated into the model when identified. Early models also inform the placement and necessity of additional mass components required to reproduce the observed lensing configuration.  Throughout the modeling for \clustershort, we identified additional substructure within both the primary and secondary arcs, but no other model-predicted counter images could be securely identified.

In the early stages, models are optimized in the source-plane, where observed image positions are ray-traced back to the source-plane and the scatter in source positions is minimized. In this approach, the $\chi^2$ is computed from the dispersion between the predicted source positions corresponding to each set of multiple images. Source-plane optimization is relatively fast and well suited for initial exploration of parameter space, but does not directly minimize residuals in the image-plane. As the modeling process progresses, we transition to the more computationally-intensive image-plane optimization, in which model predicted image positions are directly compared to the observed arc locations. In this case the $\chi^2$ is defined from the positional offsets between the predicted and observed image positions. The models presented here are obtained under image-plane optimization using \lenstool\ version 8.6.4.

We assume the same positional error for all image constraints. This is much higher than the JWST astrometry to account for small deviations in lensing deflections from mass substructures in the lens or line of sight that are not accounted for in the lens model. While the assumed positional uncertainty has a moderate impact on the posterior distribution, different values of this parameter result in similar best-fit models due to where it enters the $\chi^2$ calculation. In order to ensure the robustness of reported uncertainties in our final model, we iteratively identified the appropriate positional uncertainty that produces a reduced $\chi^2$ of order unity ($\chi^2_\nu \sim 1$). The positional error parameter in the final model is $0\farcs1$.
Our final model contains 93 degrees of freedom and achieves a $\chi^2$ of 108.0. It is constrained by a total of 88 multiple images of clumps in four background sources, and has an image-plane scatter of $\mathrm{RMS}_\mathrm{i}=0\farcs11$.

\subsubsection{Lens Plane Mass Components}
The lensing mass distribution of \clustershort\ is modeled as the combination of a smooth cluster scale component and galaxy scale perturbations. We model all mass components as pseudo-isothermal elliptical mass distributions (dPIE, also referred to in the \lenstool\ literature as PIEMD; \citealt{Limousin2005}), with each halo described by seven parameters: the centroid position $(x,y)$, ellipticity $e = (a^2 - b^2)/(a^2 + b^2)$ (where $a$ and $b$ are the semi-major and semi-minor axes), position angle $\theta$, fiducial velocity dispersion $\sigma$, core radius $r_{\mathrm{core}}$, and cut radius $r_{\mathrm{cut}}$. The halos are listed in \autoref{tab:model}, and described below.

\textit{Cluster scale halo---} The smooth component of the cluster mass distribution is modeled with a dPIE halo whose centroid is confined by priors to remain within $5\arcsec$ of the BCG. We allow all parameters to vary except the cut radius, which is fixed to $r_{\mathrm{cut}} = 1500$ kpc, as it lies well beyond the strong lensing region and cannot be constrained by strong lensing alone. We refer to this component as halo H1 throughout the remainder of this work. 

\textit{BCG---} A dPIE halo centered on the BCG allows the inner mass slope to vary independently of the large-scale cluster halo or other cluster-member galaxies, and improves reconstruction of lensing features sensitive to the central mass distribution. All parameters of this component are free except for its centroid $(x,y)$, which is fixed to the BCG position. We refer to this component as halo H2.

\textit{Galaxy scale halos---} Each galaxy in the cluster member catalog is assigned a dPIE halo under the assumption of a mass-to-light correlation. The halo parameters $\sigma_0$, $r_\mathrm{core}$, and $r_\mathrm{cut}$ are scaled with F105W magnitudes following the relations in \citet{Jullo2007}:
\begin{equation}
    \sigma_0=\sigma_0^\star \left( \frac{L}{L^\star}\right)^{1/4}
    \label{eq:sigma}
\end{equation}
\begin{equation}
    r_\mathrm{core}=r_\mathrm{core}^\star \left( \frac{L}{L^\star}\right)^{1/2}
    \label{eq:rcore}
\end{equation}
\begin{equation}
   r_\mathrm{cut}=r_\mathrm{cut}^\star \left( \frac{L}{L^\star}\right)^{1/2}
   \label{eq:rcut}
\end{equation}
Where $\sigma^\star_0$, $r^\star_\mathrm{core}$, and $r^\star_\mathrm{cut}$ represent parameters for an $L^\star$ galaxy. We determine the apparent F105W magnitude of an $L^\star$ galaxy at \clustz\ to be $m^\star = 19.9$. This is computed by convolving a synthetic spectrum of an $L^\star$ elliptical galaxy, evolved to the cluster redshift, with the WFC3/IR F105W filter response. The parameters $\sigma^\star$ and $r_\mathrm{cut}^\star$ are freed in the optimization, and we fix $r_{\mathrm{core}}^\star = 0.15$ kpc as it is weakly constrained by the lensing data.

\textit{Individually modeled galaxies---} We treat several galaxies independently rather than following the red-sequence scaling relations. These galaxies are removed from the catalog to avoid double counting their contribution to the lensing potential.

Similarly to \citet{Johnson2017}, we include in the model a circular halo for a galaxy that appears embedded within Arc 1.3. The nearby arc contaminates its photometry and affects its measured colors and structural parameters, making their measurements unreliable. We therefore model this galaxy as a circular dPIE halo with only the velocity dispersion allowed to vary. The core and cut radii are fixed to values derived from the cluster-member scaling relations, as determined in earlier models where this galaxy was manually included in the galaxy catalog. When these parameters were later freed as a test, we found that they were poorly constrained and consistently converged to those same values. This component is labeled H3.


A blue, spectroscopically confirmed cluster member at $\mathrm{R.A.}=\text{11:10:18.4733}$, $\mathrm{decl.}=\text{+64:59:58.567}$  does not lie on the red sequence and exhibits morphology characteristic of a ``jellyfish" galaxy \citep{Johnson2017}; displaying trails of knotted star formation produced as it moves through the hot intracluster medium and is stripped of its cold gas \citep{Ebeling2014}. As it does not follow the red-sequence scaling relations, we model this galaxy as an independent dPIE halo. The velocity dispersion and cut radius are allowed to vary, while the ellipticity and position angle are fixed to values measured from the stellar light distribution in WFC3/IR F105W. The core radius is fixed to $r_{\mathrm{core}} = 0.1$ kpc, as it is not well constrained. 
Models that allow wide priors on the free parameters of this ``jellyfish" galaxy result in the production of an additional counter image of system 3 that is not observed in the data. In these models, the halo converges to a higher mass normalization, extending the primary critical curve to this region and predicting an extra image. To prevent the formation of an unobserved image of system 3, we impose an upper limit on the velocity dispersion of this halo, restricting $\sigma_0 \leq115\,\mathrm{km\,s^{-1}}$. This component is labeled as H4.

We further include a halo on a cluster member located at $\mathrm{R.A.}=\text{11:10:21.1865}$, $\mathrm{decl.}=\text{+64:59:42.382}$, south of Arc 1.2. Residual positional offsets between observed and model-predicted images in this region indicate the need for a localized mass perturbation. For this halo, only the velocity dispersion and cut radius are allowed to vary; the remaining parameters are fixed to values informed by the galaxy’s stellar light distribution and initial scaling relations. We label this component as H5.

\textit{External shear---} Finally, we include an external shear component in the lens model to account for tidal gravitational perturbations from mass structures outside the modeled field of view. Such perturbations can arise from nearby groups or clusters along the line of sight and can introduce a large-scale distortion to the lensing potential. In the case of \clustershort, a galaxy cluster at $z=0.346$ located at $\mathrm{R.A}=\text{11:10:35.4912}$, $\mathrm{decl.}=\text{+64:58:55.905}$ approximately $2'$ southeast of the lens center was identified in SDSS \citep[e.g., the GMBCG Galaxy Cluster Catalog;][]{Hao2010}. In addition, wide-field weak-lensing analysis of \clustershort\ by \citet{Oguri2012} reveals evidence for large-scale mass structure surrounding the system.

\begin{deluxetable*}{lccccccccc}
\tablecaption{Best-fit Lens Model Parameters\label{tab:model}}
\setlength{\tabcolsep}{3pt}
\tablehead{
\colhead{Component} &
\multicolumn{7}{c}{dPIE halo parameters} &
\multicolumn{2}{c}{External shear} \\
\colhead{ } &
\colhead{$x$ ($''$)} &
\colhead{$y$ ($''$)} &
\colhead{$e$} &
\colhead{$\theta$ (deg)} &
\colhead{$\sigma$ (km s$^{-1}$)} &
\colhead{$r_{\rm core}$ (kpc)} &
\colhead{$r_{\rm cut}$ (kpc)} &
\colhead{$\gamma$} &
\colhead{$\theta_\gamma$ (deg)}
}
\startdata
H1 (Cluster) 
& $0.70^{+0.99}_{-1.16}$ & $-0.72^{+0.23}_{-0.25}$ & $0.791^{+0.008}_{-0.025}$ & $163.02^{+0.93}_{-0.83}$ & $737^{+41}_{-14}$ & $36.2^{+11.1}_{-2.5}$ & [1500] & --- & --- \\
H2 (BCG) & [0] & [0] & $0.446^{+0.052}_{-0.063}$ & $-58.6^{+9.9}_{-1.1}$ & $395^{+29}_{-16}$ & $5.4^{+2.2}_{-1.1}$ & $148.78^{+0.31}_{-51.23}$ & --- & --- \\
H3 (Embedded Galaxy) & [11.61] & [10.44] & [0] & [0] & $62.3^{+3.8}_{-5.7}$ & [0.03] & [25.36] & --- & --- \\
H4 (Jellyfish) & [4.86] & [10.70] & [0.25] & [-15.01] & $110.2^{+4.4}_{-35.0}$ & [0.1] & $12.4^{+2.5}_{-8.0}$ & --- & --- \\
H5 (South Perturber) & [22.07] & [-5.49] & [0.01] & [73.86] & $201^{+44}_{-12}$ & [0.06] & $56.7^{+2.8}_{-29.1}$ & --- & --- \\
$L^\star$ galaxy (scaling) & --- & --- & --- & --- & $204^{+28}_{-80}$ & [0.15] & $17.9^{+72.7}_{-6.3}$ & --- & --- \\
External shear & --- & --- & --- & --- & --- & --- & --- & $0.091^{+0.044}_{-0.010}$ & $137.7^{+10.6}_{-6.1}$ \\
\enddata
\tablecomments{Parameters of the best-fit strong lensing model of \clustershort. All coordinates are measured in arcseconds relative to the center of the BCG, at [R.A., Decl.] = [167.573775, 64.996630]. Error bars correspond to the 95\% confidence interval inferred from the full MCMC posterior. Values in square brackets are for parameters that were not optimized.}
\end{deluxetable*}

\subsubsection{Testing Modeling Choices}
\label{subsubsec:bic_tests}
To evaluate whether the inclusion of halo H5 and the external shear are statistically warranted, we compare models with and without these components using the Bayesian Information Criterion (BIC). BIC provides a quantitative framework for comparing models with differing numbers of free parameters. This mitigates overfitting by penalizing improvements in model likelihood that arise solely from the addition of unconstrained or weakly constrained parameters. We note that BIC is not well suited for comparing fundamentally different model families, but for incremental modifications such as the addition of individual components, it provides a useful statistical test. A model with lower BIC is considered to be statistically preferred. The BIC is calculated using the following equation \citep{Schwarz1978},
\begin{equation}
    BIC=-2\ln(\mathcal{L})+k\ln(n)
    \label{eq:BIC}
\end{equation}
where $\mathcal{L}$ is the likelihood, $k$ is the number of free parameters, and $n$ is the sample size. In our analysis, $n$ is taken to be the number of independent positional constraints, defined as
\begin{equation}
    n=2N_\mathrm{img}-2N_\mathrm{src}
\end{equation}
where $N_{\mathrm{img}}$ is the total number of observed multiple images of unique source clumps and $N_{\mathrm{src}}$ is the number of unique source clumps (see \citealt{Lorah2019} for discussion on sample size versus constraints). In this framework, $N_{\mathrm{src}}$ corresponds to the number of independently solved source-plane coordinates. For our model, $N_{\mathrm{src}} = 30$ and $N_{\mathrm{img}} = 88$, which yields $n = 116$ independent positional constraints.

We compute BIC values for four models: (1) a model including both external shear and H5, (2) a model including H5 but no shear, (3) a model including shear but not H5, and (4) a model including neither component. The resulting BIC values are listed in \autoref{tab:BIC}. We find that the lowest BIC is obtained for the model including both the H5 and the external shear, indicating that the improvement in likelihood outweighs the penalty associated with the additional free parameters. This supports the statistical necessity of both components. 

\begin{deluxetable}{cccccccc}
\tablecaption{Model BIC Comparisons \label{tab:BIC}}
\tablehead{
\colhead{Model} &
\colhead{$n$} &
\colhead{$k$} &
\colhead{dof} &
\colhead{$\ln\mathcal{L}$} &
\colhead{$\chi^2$} &
\colhead{$\mathrm{RMS}$} &
\colhead{BIC}
}
\startdata
(1)&  116&  23& 93 & 189.5& 108.0&$0\farcs11$ & -331.6\\
(2)&  116&  21& 95 & 57.1& 372.8&$0\farcs21$ & -70.9\\
(3)&  116&  21 & 95 & 151.1& 184.8&$0\farcs14$ & -258.9\\
(4)&  116&  19& 97 & -24.3& 535.7&$0\farcs25$ & 87.8\\
 \hline
\enddata
See \autoref{eq:BIC} for definitions of the parameters.
\tablenotetext{}{(1) Model with shear and H5.}
\tablenotetext{}{(2) Model without shear.}
\tablenotetext{}{(3) Model without H5.}
\tablenotetext{}{(4) Model without shear and H5.}
\end{deluxetable}

\subsection{Magnification Calculation}
\label{sec:tot_mag_methods}

One of the outputs of a lens model is a redshift-dependent magnification matrix, which maps the lensing magnification at any $x$, $y$ position in the image-plane. The magnification of \textit{extended} arcs typically varies along the arc, reaching very high values near the critical curve. The total areal magnification of such arcs can be calculated as the ratio between its total image-plane area and its total model-predicted source-plane area. Below, we describe how the boundaries of each arc were defined, and how the source-plane area was calculated, for a given lens model.

We define the area over which magnification is measured using the F182M filter, which  provides strong contrast between the arc emission and the surrounding background due to the strong $[\textrm{O}\textrm{III}]$ emission, clearly revealing the full spatial extent of the arcs. Background statistics are estimated using sigma-clipped measurements. Pixels exceeding $3\sigma$ above the background are first used to define a high-confidence core of the arc emission. The masks are then expanded by including connected pixels exceeding $1.5\sigma$ above the background, allowing us to capture the full spatial extent of the arcs while minimizing the inclusion of isolated noise fluctuations.

We verified that the resulting magnifications are not sensitive to the exact threshold choice by varying the expansion threshold within a reasonable range (e.g., $1.5$–$2\sigma$), finding negligible impact on the measured magnifications. The resulting masks isolate the emission associated with the three lensed arcs and define the regions over which magnification is measured.

To ensure that each mask corresponds to a single lensed image, we restricted the growth of the arc regions using the critical curve predicted by the lens model for a source at \galz. The critical curve marks the location in the image-plane where the lensing magnification formally diverges and separates regions of different image parity. Pixels were included in a given arc mask only if they lay on the same side of the critical curve, preventing the segmentation from crossing into neighboring lensed images. The critical curve therefore acts as a natural boundary line between distinct images of the background source. For each model used from the posterior distribution, the critical curve corresponding to that model was computed, ensuring consistency between the magnification field and the segmentation of the arc boundaries.  The masks corresponding to the best-fit model are shown as gray regions in \autoref{fig:magnif}.

For each arc, the total areal magnification was computed from the ratio of the image-plane area ($A_{\rm img}$) to the model predicted source-plane area ($A_{\rm src}$), 
\begin{equation}
\mu_{\rm arc} = \frac{A_{\rm img}}{A_{\rm src}}.
\label{eq:muarc}
\end{equation}
The image-plane area was obtained by summing the pixel area, $A_{\rm pix}$, across the arc mask. The source-plane area was computed using the local magnification values from the model,
\begin{equation}
A_{\rm src} = \sum_i \frac{A_{{\rm pix,}i}}{\mu_i},
\label{eq:Asrc}
\end{equation}
where $A_{{\rm pix,}i}$ is the image area at pixel $i$ and $\mu_i$ is the magnification at pixel $i$. Combining \autoref{eq:muarc} and \autoref{eq:Asrc} therefore yields the magnification of each arc.

The total areal magnification of the system, $\mu_\mathrm{tot}$, was then computed as the sum of the magnifications of the three arcs. Magnification uncertainties were derived by repeating the process for a set of models drawn from the posterior distribution, producing a distribution of magnification values for each arc and the total magnification.

Individual clump magnifications are estimated directly from the magnification maps, which were computed for the source redshift and $0\farcs03$ px$^{-1}$ resolution.  
The magnification assigned to each clump corresponds to the value of the magnification map at the coordinate of the clump centroid, computed with linear 2-d interpolation of the magnification map.
Uncertainties on clump magnifications are derived by evaluating the magnification at the same location across the ensemble of models sampled from the posteriors.
The resulting magnifications of individual clumps and the total galaxy are presented in \autoref{sec:results}, and tabulated in Tables~\ref{tab:constraints} and \ref{tab:magnif}.

\subsection{Source-Plane Properties}
\label{sec:source_plane_methods}
We produce several source-plane products to facilitate the analysis that will be discussed later in the paper.

Source-plane projections are produced using \lenstool's \texttt{cleanlens} mode. In this process, image-plane pixels belonging to each arc are ray-traced through the lens equation and resampled onto a finer grid in the source-plane at $z=2.481$, producing three separate reconstructions of the same source.

To estimate source-plane width ($W$), length ($L$) and $W/L$ ratio, we first ray-trace the arc masks described in \autoref{sec:tot_mag_methods} using the deflection maps produced by \lenstool. We then fit an ellipse to each projected mask, and define $W$ and $L$ as the minor and major axes of the ellipse, respectively.

We also calculate the source-plane flux, $F_{\nu,\mathrm{src}}$, from the image-plane surface brightness and local magnification. The JWST imaging provides the surface brightness $I_\nu$ in units of $\mathrm{MJy\,sr^{-1}}$. Gravitational lensing preserves surface brightness, while the source-plane solid angle corresponding to each image-plane pixel is $\Omega_{\mathrm{src},i}=\Omega_{\mathrm{img}}/|\mu_i|$, where $\Omega_{\mathrm{img}}$ is the image-plane pixel solid angle and $\mu_i$ is the local magnification. The source-plane spectral flux density is therefore
\begin{equation}
    F_{\nu,\mathrm{src}}
    =
    \sum_{i=1}^{N}
    I_{\nu,\mathrm{img},i}
    \frac{\Omega_{\mathrm{img}}}{|\mu_i|},
\end{equation}
where the sum extends over the $N$ pixels within the corresponding arc mask. This yields three separate estimates of the source-plane spectral flux density of \galaxyshort.

The geometric and flux density measurements are repeated for every model drawn from the posterior distribution, ensuring consistency between the deflection field and the arc boundaries. 

\section{Results}\label{sec:results}
We identify the best-fit lens model as the realization that minimizes the $\chi^2$ during the \lenstool\ image-plane optimization, as described in \autoref{sec:lensmodel}. Uncertainties on the optimized model parameters are obtained directly from the \lenstool\ MCMC sampling. For each parameter, we report the best-fit value together with asymmetric uncertainties defined by the 95\% confidence interval of the full posterior distribution.

For quantities derived from the lens model (e.g., enclosed mass and magnification), uncertainties are estimated from 300 models randomly drawn from the posterior distribution, including the best-fit model. We compute the derived quantities from each realization, and report the best-fit value with uncertainties defined by the 95\% confidence interval of the posterior distribution. In this section, we examine the best-fit model, and present lens model results including cluster lens mass, magnifications, and source-plane properties. We provide time delay predictions in \autoref{sec:appendix}. The results are discussed in \autoref{sec:discussion}.

\subsection{Best-Fit Lens Model}

The final model includes the cluster-scale halo (H1), the BCG-centered halo (H2), the embedded galaxy halo (H3), the jellyfish galaxy halo (H4), the additional galaxy halo south of Arc 1.2 (H5), an external shear component, and contribution from cluster-member galaxies; this model corresponds to the preferred model (1) based on the BIC (see \autoref{subsubsec:bic_tests}).

The critical curve for a source at $z=2.481$ is shown in \autoref{fig:cstr}. The model reproduces the morphology and parity of the three images of the primary arc, as well as the positions and parity of the secondary arc systems.

The best-fit model has an image-plane RMS of $0\farcs11$, defined as the root-mean-square positional offset between the observed locations of the multiply imaged features and their image positions predicted by the best-fit model. The corresponding source-plane RMS is $0\farcs048$. The model has image-plane $\chi^2 = 108.0$ and 93 degrees of freedom, corresponding to $\chi^2_\nu$ of order unity. Since $\chi^2$ depends on the assumed positional uncertainties of the image constraints (taken here to be $0\farcs1$), modest changes to this value would rescale $\chi^2_\nu$ without significantly altering the underlying model quality. For this reason, the image-plane RMS provides a more robust metric for assessing the model performance. A $\chi^2_\nu$ of order unity indicates that the model reproduces the data at a level consistent with the assumed uncertainties, implying a statistically meaningful fit and posterior.

The optimized parameters for the large-scale halo and selected galaxy halos are listed in \autoref{tab:model}. The centroid of the primary cluster halo is found to lie close to the BCG position ($\Delta \rm{R.A.} \approx 0\farcs70 , \Delta \rm{decl.} \approx - 0\farcs72$), consistent with expectations for a relaxed cluster-scale potential. The large-scale dark matter halo is well described by a pseudo-isothermal elliptical mass distribution with ellipticity $e \approx 0.80$ and a velocity dispersion of $\sigma \approx 737~{\rm km~s^{-1}}$.

We further examine the external shear component introduced in the model. The best-fit model yields a position angle of $\sim140^\circ$ for the external shear\footnote{In \lenstool, the shear is measured CCW from the positive (East) x-axis \citep{Marceau2025}.}, and a shear strength of $\sim0.1$, indicating a modest but non-negligible tidal contribution. The orientation of this shear is consistent with the direction toward the foreground cluster at $z = 0.346$ \citep{Hao2010}, supporting the interpretation that this structure contributes additional lensing potential along this line of sight.

The model-predicted redshifts for the secondary arc systems are $z_{\mathrm{model},2}=2.96^{+0.46}_{-0.28}$, $z_{\mathrm{model},3}=3.07^{+0.22}_{-0.13}$, $z_{\mathrm{model},4}=4.80^{+0.16}_{-1.57}$. These values lie within the broad ranges of their photometric redshift posteriors. However, given the large extent of the photometric redshift intervals and sizable uncertainties on the model-predicted redshifts, particularly system 4, this agreement does not provide a strong constraint on the secondary arc redshifts.

\subsection{Mass of the Lensing Cluster \clustername}
\label{sec:mass}

We measure the projected mass distribution at the core of the lensing cluster using the ensemble of models drawn from the \lenstool posterior distribution. We compute the mass enclosed within a projected radius of 250~kpc, as well as the mass enclosed by the primary critical curve for a source at $z=2.481$. We derive the effective Einstein radius from the area enclosed by the critical curve,
\begin{equation}
    \theta_E=\sqrt{A(<\rm{crit})/\pi}.
    \label{eq:einstein_rad}
\end{equation}
We find a projected mass of \masskpc. The area and mass enclosed by the primary critical curve are \areacrit\ and \masscrit, respectively. The resulting Einstein radius is \einstrad.

\subsection{Magnification of \galaxyname}

To determine the total magnification of the primary arc and its uncertainties, we apply the procedure described in \autoref{sec:tot_mag_methods} to the ensemble of lens models drawn from the posterior distribution. For each realization, we compute the integrated magnification of each individual arc and the total magnification, $\mu_\mathrm{tot}$, defined as the sum of the magnifications across all arc components.
The best-fit magnification and statistical uncertainties are tabulated in the first row of  \autoref{tab:magnif}.

Across the identified clumps in the best-fit model, the local magnification ranges from $\sim6-15$ in arc 1.1, $\sim5-19$ in arc 1.2, and $\sim4-12$ in arc 1.3. Individual clump magnifications are listed in \autoref{tab:constraints}.

\begin{figure*}
\centering
    \includegraphics[width=1\linewidth]{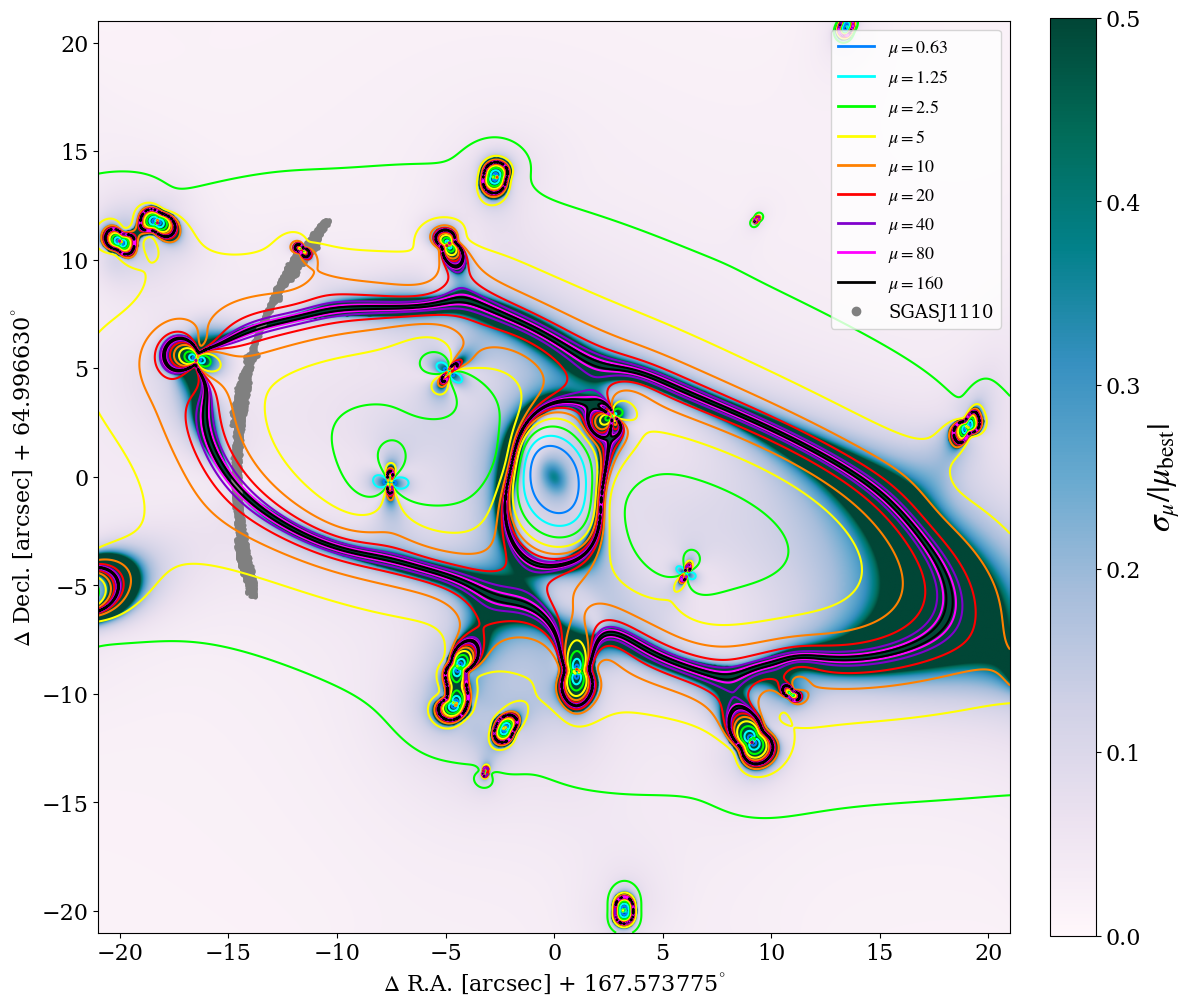}
    \caption{Magnification and fractional magnification uncertainty across the image-plane for the best-fit lens model. The background color map shows the fractional uncertainty in magnification, computed from the standard deviation of magnification values across 300 lens models randomly drawn from the posterior distribution and normalized by the magnification of the best-fit model. The color scale is truncated at 0.5 for visualization. To prevent extreme magnifications near the critical curve from artificially suppressing the fractional uncertainty, magnification values exceeding $\mu=500$ are capped when computing the fractional uncertainty. Colored contours show constant magnification levels predicted by the best-fit model, and the gray region marks the location of the primary arc system.} 
    \label{fig:magnif}
\end{figure*}

\autoref{fig:magnif} shows the spatial distribution of magnification and its associated uncertainty across the image-plane. The colored background represents the fractional uncertainty in magnification, defined as $\sigma_\mu/|\mu_{\rm best}|$, where $\sigma_\mu$ is the standard deviation of magnification values across the 300 models sampled from posterior distribution and $\mu_{\rm best}$ is the magnification from the best-fit model. To avoid artificially suppressing uncertainties in regions of extremely high magnification, values exceeding $\mu = 500$ are capped at this threshold when computing the fractional uncertainty. The color scale is truncated at a fractional uncertainty of 0.5 for visualization purposes. Contours indicate constant magnification levels from the best-fit model. Regions of higher fractional uncertainty are primarily located near the critical curves and in areas of the field with fewer lensing constraints, particularly toward the western side. Notably, the highly constrained regions, specifically along the giant arc, have among the lowest fractional magnification uncertainty range of $\sigma_\mu/|\mu_{\rm best}|=0.03-0.09$.

\begin{deluxetable}{ccccc}
\tabletypesize{\scriptsize}
\tablecaption{Arc and Total Magnification Values\label{tab:magnif}}
\tablehead{
\colhead{Model} &
\colhead{$\mu_{1.1}$} &
\colhead{$\mu_{1.2}$} &
\colhead{$\mu_{1.3}$} &
\colhead{$\mu_{\rm tot}$}
}
\startdata
Full & $9.79^{+1.58}_{-0.53}$ & $7.66^{+1.17}_{-0.57}$ & $6.74^{+0.67}_{-0.23}$ & $24.2^{+3.4}_{-1.2}$ \\
Diluted & $8.31^{+2.52}_{-1.21}$ & $6.70^{+1.67}_{-0.85}$ & $6.12^{+1.34}_{-0.69}$ & $21.1^{+5.3}_{-2.7}$ \\
J17-M2 & $15.7^{+4.3}_{-9.0}$ & $14.4^{+5.2}_{-8.0}$ & $12.4^{+3.9}_{-6.1}$ & $42.5^{+8.8}_{-22.5}$ \\
\enddata
\tablecomments{Full denotes the 2026 model with all constraints and spectroscopic information, Diluted denotes the reduced-constraint model under the same modeling framework as the Full model, and J17-M2 denotes Model 2 from \citet{Johnson2017}. Values correspond to the best-fit magnifications, with uncertainties derived from the 95\% posterior intervals.}
\end{deluxetable}

\subsection{Magnification of Secondary Arcs}
We compute magnifications of clumps in the secondary multiply-imaged systems, listed in \autoref{tab:constraints}. Across systems 2 and 3, the magnifications are modest, typically $\mu \sim 2-7$. System 4 exhibits significantly higher magnifications, $\mu \sim 20-30$, with high uncertainties due to its proximity to the radial critical curve. We emphasize that the tabulated values are for the best-fit redshifts and their uncertainties are underestimated since they reflect the statistical uncertainties of model parameters and do not fold in the redshift uncertainty. Spectroscopic redshifts for secondary arcs are required for a more accurate and precise magnification estimate of these sources.

\subsection{Source-Plane Projection}\label{sec:WL}

\begin{figure*}
\centering
    \includegraphics[width=1\linewidth]{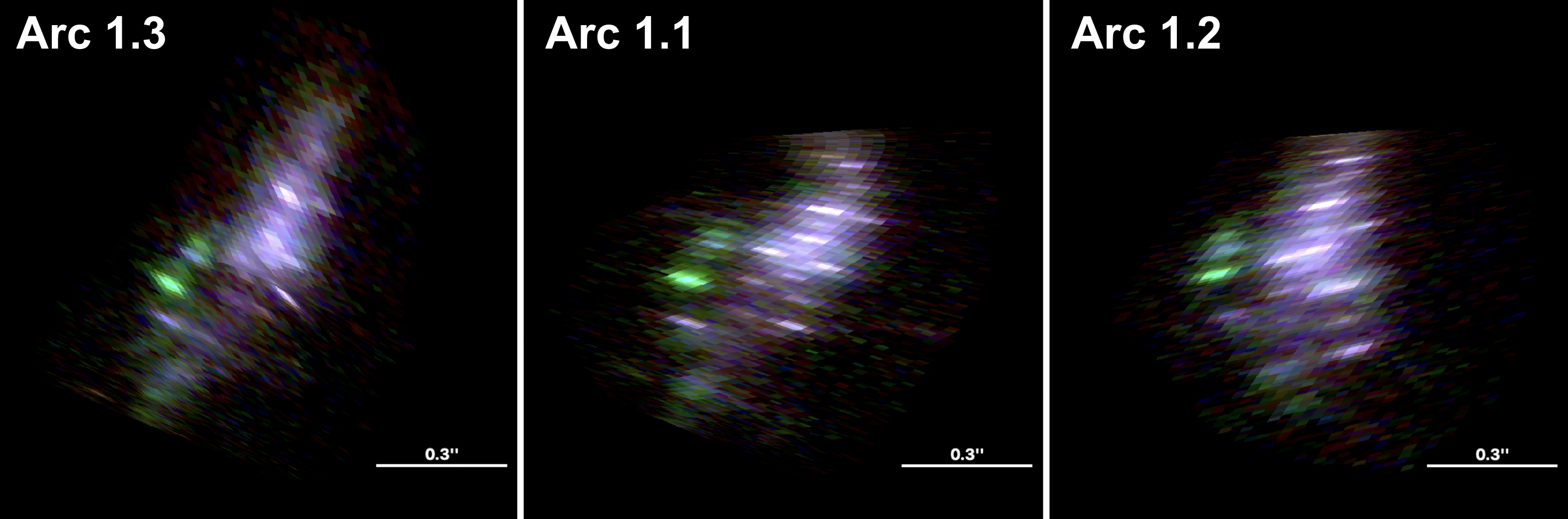}
    \caption{Source-plane projections of \galaxyshort\ obtained by ray-tracing each multiply imaged arc through the best-fit lens model. From left to right, the panels show the reconstructions derived from arcs 1.3, 1.1, and 1.2, respectively. Each reconstruction represents an independent mapping of the same background galaxy to the source-plane at \galz. The images are shown as RGB composites constructed from the F200W, F182M, and F115W filters. The consistent recovery of the overall morphology and the relative positions of the brightest clumps across the three reconstructions provides a qualitative validation of the lens model. The scale bar of $0\farcs3$  corresponds to 2.43 kpc in the source-plane.} 
    \label{fig:splane_proj}
\end{figure*}

The three source-plane reconstructions of \galaxyshort\ are shown in \autoref{fig:splane_proj}. The consistency between these reconstructions serves as a qualitative validation of the lens model. The resulting source spans $\sim7~\mathrm{kpc}$ in extent and exhibits multiple bright clumps that are recovered across all reconstructions, with slight offsets consistent with the image-plane RMS of $0\farcs11$ and source-plane RMS of $0\farcs048$.

While the quality of the model is primarily quantified through the image-plane RMS, a comparison of the source-plane properties of the three images of the primary arc provides an informative complementary assessment. \autoref{tab:source_plane_properties} lists the source-plane width, length, W/L, and source-plane spectral flux density of the three source projections, with statistical uncertainties derived from sampling the MCMC parameter space. 
To quantify the variation among the three source-plane reconstructions, we calculate the fractional scatter in the properties recovered from each arc using the best-fit lens model. The fractional scatter in $W$, $L$ is $7.6\%$, $9.2\%$, respectively, indicating that the source width and length are recovered consistently at the $<10\%$ level. The ellipticity of the source, characterized by $W/L$, is less consistent among reconstructions, with a fractional scatter of $18.8\%$. The source-plane flux, $F_{\nu,\mathrm{src}}$, is recovered remarkably consistently, with $4.1\%$ scatter. We discuss the inferred systematic uncertainties in \autoref{subsec:systematic_uncert}.

\begin{deluxetable}{ccccc}
\tabletypesize{\scriptsize}
\tablecaption{Source-Plane Properties of \galaxyshort \label{tab:source_plane_properties}}
\tablehead{
\colhead{Arc} &
\colhead{$W\,[\mathrm{kpc}]$} &
\colhead{$L\,[\mathrm{kpc}]$} &
\colhead{$W/L$} &
\colhead{$F_{\nu,\mathrm{src}}\,[\mu\mathrm{Jy}]$}
}
\startdata
1.1 & $3.30^{+0.06}_{-0.22}$ & $7.60^{+0.33}_{-0.74}$ & $0.43\pm0.03$ & $1.06^{+0.07}_{-0.16}$ \\
1.2 & $3.55^{+0.10}_{-0.24}$ & $6.13^{+0.58}_{-0.69}$ & $0.58^{+0.07}_{-0.06}$ & $1.05^{+0.09}_{-0.14}$ \\
1.3 & $3.06^{+0.11}_{-0.25}$ & $7.22^{+0.34}_{-0.51}$ & $0.42\pm0.03$ & $0.98^{+0.05}_{-0.09}$ \\
Combined & $3.22^{+0.39}_{-0.38}$ & $7.10^{+0.77}_{-1.51}$ & $0.44^{+0.19}_{-0.04}$ &  $0.99^{+0.13}_{-0.09}$ \\
\enddata
\tablecomments{Source-plane size and spectral flux densities of \galaxyshort\ inferred from arcs 1.1, 1.2, and 1.3. Spectral flux densities are evaluated in the NIRCam F182M imaging. For each individual arc, the central values correspond to the best-fit lens model, with uncertainties derived from the 95\% confidence intervals of the MCMC posterior. The combined central values represent the median of the posterior samples from all three arcs, with uncertainties spanning the corresponding 95\% posterior interval.
\label{tab:source_plane_properties}}
\end{deluxetable}

\section{Discussion and Implications}\label{sec:discussion}

In this section, we estimate the systematic uncertainties, compare the results to previous work, and examine their implications.

\subsection{Assessment of Systematic Uncertainties}
\label{subsec:systematic_uncert}

It is well known that statistical uncertainties inferred from, e.g., MCMC sampling of the parameter space, underestimate the total lens modeling uncertainty; systematic uncertainties come from differences between modeling algorithms, modeler choices, correlated and uncorrelated structure not accounted for by the model, and the availability and distribution of constraints \citep{Treu2010,Zitrin2015,Rodney2015,Johnson&Sharon2016,Acebron2017,Meneghetti2017}. Degeneracies in the absolute mass normalization, such as the mass-sheet degeneracy, allow different mass distributions to reproduce the same observed image positions and relative distortions while predicting different absolute magnifications \citep{Schneider_Seitz_1995}. Independent magnification tests using lensed Type~Ia supernovae further illustrate the limitations of statistical uncertainties \citep{Rodney2015}.

\citet{Meneghetti2017} compared parametric, free-form, and hybrid reconstructions of two simulated Hubble Frontier Fields-like clusters for which the true mass and magnification fields were known. They found systematic magnification errors of approximately 10\% at $\mu_\mathrm{true}=3$, increasing to approximately 30\% at $\mu_\mathrm{true}\sim10$, and becoming larger at higher magnification. This assessment can serve as a useful reference point for systematic uncertainties in similar lensing systems. For the \galaxyshort\ clumps, which have magnification range of 4-19, this would imply a systematic uncertainty of order 13-57\%, assuming a linear relation between magnification and uncertainty. However, in \citet{Meneghetti2017}, the simulated arcs did not contain the same density of JWST resolved clumps that is present in \galaxyshort.

To obtain a systematic uncertainty estimate more closely matched to our data, we compare two independent lens models of the lensing cluster RCS2~032727$-$132623 (hereafter RCS 0327) presented by \citet{Cerny2026}. RCS 0327 lenses a galaxy at $z=1.701$ into a $\sim38\arcsec$ long giant arc, similar to \galaxyshort, containing a dense sample of 42 distinct JWST-resolved clumps producing 112 individual clump images. \citet{Cerny2026} modeled this cluster with both \lenstool\ and the hybrid free-form code \wslap\ \citep{Diego2007}. The two reconstructions agree in enclosed mass to within approximately 4\% at 200 and 500 kpc.

We evaluate the best-fit magnification maps from the two models at every clump position for the galaxy at $z=1.701$ and define the fractional difference as $\Delta f=|\mu_\lenstool-\mu_\wslap|/|\mu_\lenstool|$. Across all matched clump images, the median difference is $6.7\%$, more than half of the clumps have fractional difference $<10\%$, and $90\%$ of the clumps have a fractional difference of $<40\%$. 
Thus, the two methods predict similar magnifications at most clump positions despite their different modeling approaches and constraint sets. Large differences are found primarily near the model critical curves, with a small number of these clumps exceeding a fractional difference of more than $40\%$ and only a handful of clumps placed within $0\farcs5$ from the critical curve reach $100\%$ or above. This comparison indicates that method-dependent magnification uncertainties are modest across most of the arc but can become dominant for clumps lying close to a critical curve. Because the models also use slightly different constraint sets, we interpret the results of this test more generally as an empirical estimate of systematic uncertainty.

We adopt a representative fractional systematic uncertainty of  $\sim7\%$ for the magnifications of the \galaxyshort\ clumps, based on the median difference between the two independent lens models of RCS0327. While systematic uncertainties can reach tens of percent or more for clumps very close ($\lesssim1-2''$) to the critical curves, this is not a concern for most of the identified clumps in \galaxyshort.

Another handle on our modeling uncertainties may be obtained from the source projections presented in \autoref{sec:WL}. The scatter among the best-fit source properties inferred from arcs 1.1, 1.2, and 1.3 measures how consistently the same source is recovered through three different regions of the modeled lensing potential. 
We find that the physical dimensions of the three source projections are consistent to within approximately $10\%$, comparable to the $\sim2-11\%$ statistical uncertainty on the source width and length (\autoref{tab:source_plane_properties}); the $W/L$ ratio has a larger reconstruction-to-reconstruction variation of $18.8\%$, mostly driven by image 1.2. 
While the overall agreement between reconstructions remains strong, the source-plane curvature is not perfectly consistent across all projections. In particular, the curvature of the source projection of arc 1.2 is slightly different from those of arcs 1.1 and 1.3, which is also apparent in its larger $W/L$. This discrepancy is modest and likely reflects small-scale structure or higher-order perturbations in the lensing potential that are not captured by the current model. 
The inferred source-plane flux is particularly consistent among the three arcs, only $4.1\%$, smaller than the uncertainties in any of the geometric properties, suggesting that the inferred intrinsic spectral flux density is comparatively insensitive to residual differences in the lens mapping.

This test complements the MCMC-derived statistical uncertainties to provide an empirical estimate of the uncertainty in the inferred global source-plane properties, indicating an estimated 10\% systematic uncertainty on the size of the galaxy, that may be added to the statistical $W$ and $L$ uncertainties in quadrature.

\subsection{Comparison to Previous Work}
\subsubsection{Mass Distribution}\label{subsec:mass}

The global lensing properties of the cluster are consistent with \citet{Johnson2017}. The Einstein radius and critical curve area are in close agreement, as is the mass enclosed within the critical curve (\autoref{sec:mass}). This is consistent with expectations from strong lensing, as strong lensing constraints robustly fix the enclosed mass within the region directly probed by the arcs \citep[e.g.,][]{Remolina2021}, which for \clustershort, is approximately $15''$ from the BCG.

However, differences emerge at larger radii. The projected mass within 250~kpc ($35\farcs87$) is lower than reported by \citet{Johnson2017}, reflecting a systematic uncertainty associated with modeling choices and the treatment of mass outside the well-constrained region. While strong lensing provides precise constraints on the enclosed mass within the critical curve, the mass distribution at larger radii is effectively extrapolated from the inner profile (in parametric models) and is therefore more sensitive to assumptions about the underlying mass distribution.

External structures can contribute to the lensing potential in the cluster core and influence the observed image configurations \citep[e.g.,][]{Mahler2018}. Our model accounts for such contributions through an external shear component, whereas \citet{Johnson2017} employ a hybrid parametric and grid-based approach to capture similar effects. Because their model confines mass components to a limited spatial region within $40''$ of the BCG, contributions from more distant structures may be redistributed into intermediate radii, potentially increasing the inferred mass within 250~kpc. In contrast, our model captures the influence of external structures through a shear component without explicitly assigning mass to them. 

This comparison highlights that the statistical uncertainties in lens modeling underestimate the true uncertainty at projected radii beyond the strong lensing regime, which is dominated by systematic effects related to model assumptions and the treatment of external mass components \citep[e.g.,][]{Zitrin2015,Meneghetti2017}.

\subsubsection{Improved Positional Accuracy and Magnification Estimates}\label{sec:pos_acc}

Our JWST based lens model achieves image- and source-plane RMS values of $\mathrm{RMS_i}=0\farcs11$, which, naively, appears similar to the best-fit RMS values of \citet{Johnson2017}, $\mathrm{RMS_i}=0\farcs10 - 0\farcs12$. However, these models were evaluated for a different number of constraints, and therefore their RMS values are not directly comparable. Similarly, a comparison of the magnification prediction of the new model and the 2017 models implies that they are consistent within the broad uncertainties in \citet{Johnson2017}. However, the magnifications and uncertainties are not calculated in the same way and therefore a meaningful comparison requires additional analysis. To this end, we obtained the \lenstool\ files from \cite{Johnson2017} through private communication, verified that the values reported in the literature can be recovered from the model files, and analyzed the \cite{Johnson2017} model performance and predictions directly.

To assess the improvement in RMS we perform two complementary tests. First, to compare the performance of the models against the same set of arcs, we hold fixed the eight best-fit lens models from \citet{Johnson2017} and evaluate how well they reproduce our expanded constraint catalog and updated redshift information. We exclude system~4 because that source was not identified in the earlier analysis. When evaluated against the common expanded catalog, the eight \citet{Johnson2017} models span  $\mathrm{RMS_i}=0\farcs22-0\farcs41$ and $\mathrm{RMS_s}=0\farcs077-0\farcs169$, respectively. These RMS values are larger by factors of $2.0-3.7\times$ and $1.6-3.5\times$, from our full model, implying that the new model provides a significantly higher precision. 

Since the improvement in RMS could be due to the availability of constraints or the different modeling methodology, or both, we introduce a second test. We optimize a ``reduced constraints'' model using our parametric framework but use as constraints only the image identifications and redshift information that were available to \citet{Johnson2017}. After optimization, this model is held fixed and evaluated against the same expanded catalog without reoptimizing its mass parameters. The reduced-constraint model yields $\mathrm{RMS_i}=0\farcs19$ and $\mathrm{RMS_s}=0\farcs060$.  
Because the reduced constraint and full models employ the same parametric framework, their comparison isolates the effect of the expanded constraint set. From these two tests, we conclude that the high density of lensing constraints identified in the JWST imaging of \galaxyshort\ leads to a reduction of image-plane RMS by a factor of $1.7\times$ and the source-plane RMS by a factor of $1.25\times$.

Next, we compare the magnification predictions of the new model to the previous work. The total magnification we measure for \galaxyshort, \magtot, appears to be consistent with the value of $\mu_{\rm tot}=28\pm8$ reported by \citet{Johnson2017}. However, as noted in \autoref{subsec:prev_work}, the magnification reported in \citet{Johnson2017} is the median and median absolute deviation of the best-fit magnification values of eight different models. Our uncertainties instead describe the 95\% posterior interval obtained by sampling the parameter space of a single model. In addition, the two papers may have used different masks in the magnification calculation. To facilitate a consistent comparison, we apply the magnification methodology described in \autoref{sec:tot_mag_methods} to J17-M2 (the lowest-RMS model from \citealt{Johnson2017} and the model adopted for their subsequent source-plane analyses). The arc magnification values for our models and J17-M2 are listed in \autoref{tab:magnif}. While the best-fit total magnification from J17-M2 is $1.76\times$ higher than ours, its broad 95\% posterior interval encompasses our measurement. The same is true for the three individual arcs. Notably, the fractional uncertainties (defined as the full width of the 95\% confidence interval divided by the best-fit value) reduce from $\sim81-92\%$ in J17-M2 to $\sim 13-23\%$ in the new model, representing a median $4.0\times$ improvement in magnification precision.

At the positions of individual clumps, the J17-M2 clump magnifications have fractional 95\% interval widths spanning $79.5\%-108.2\%$, whereas those from our model span $10.9\%-33.2\%$. A comparison of the corresponding clumps gives reductions of $2.6-8.3\times$ in magnification uncertainties, with a median improvement of $4.0\times$.

Similarly to the RMS analysis, we attempt to isolate the impact of the availability of constraints on the improved magnification uncertainties using our ``diluted'' model. The magnifications predicted by this model are consistent with those from our full model, but as expected, their posterior intervals are broader. The full model has $1.7-2.5\times$ narrower uncertainties with a median improvement of $2.1\times$ for the individual arcs, and $2.0\times$ for the total magnification. Across the matched clumps, the reductions span $1.5-2.3\times$, with a median improvement of $1.8\times$. 

We conclude that the additional JWST clumps and spectroscopic information improve the magnification precision by a factor of $2\times$. The larger improvement relative to J17-M2 likely reflects the combined effects of the additional constraints, new spectroscopic information, and differences between the parametric and hybrid mass-modeling approaches.

\subsubsection{Source Morphology}
The projected source morphology appears rounder than in previous analyses \citep{Johnson2017,Rigby2017}. Applying the same geometric measurements to arc 1.1 using J17-M2, we find $W=1.47 ^{+1.52}_{-0.23}\,\mathrm{kpc}$, $L=10.6^{+3.5}_{-3.7}\,\mathrm{kpc}$, and $W/L=0.138^{+0.222}_{-0.022}$. We carry out this comparison only for arc 1.1 because this was the image used to study the source-plane structure in \citet{Rigby2017}. Compared with our reconstruction of arc 1.1 (\autoref{tab:source_plane_properties}, \autoref{fig:splane_proj}), J17-M2 produces a smaller width and $W/L$ ratio and a larger length, resulting in a substantially more elongated source. This discrepancy likely arises from local differences in the lensing potential caused by the different constraint sets and modeling techniques. 

The \citet{Johnson2017} model did not have spectroscopic confirmation that the edge clumps D, E, and F belonged to the same source as the main arc; they therefore allowed the redshifts of these constraints to vary, effectively decoupling them from the main arc. Our spectroscopic confirmation constrains these clumps to the same source-plane as the rest of the arc. To reproduce the observed displacement between the outer-edge, mid-arc, and inner-edge clumps among the three images, our model must produce a deflection gradient that results in a sheared or skewed source-plane morphology. Decoupling D-E-F from the other clumps, effectively allowed for this displacement to be attributed to a different source redshift. The larger number of constraints provided by the JWST imaging also allows our model to sample the lensing potential and mass distribution more finely, particularly in the region surrounding the arcs.

Differences in the modeling techniques can further produce different local mass distributions. Our parametric model includes an external shear term, allowing line-of-sight structure and large-scale tidal perturbations to contribute to the lensing potential without requiring their effects to be absorbed into the cluster mass components. In contrast, the utilization of a flexible grid of mass in \citet{Johnson2017} requires such effects to be absorbed in the cluster mass components. These approaches can produce different local mass distributions near the arcs, which may contribute to the more elongated source-plane morphology recovered by J17-M2.

We note that although the overall extent of $\sim7\,\mathrm{kpc}$ appears consistent with previously reported results \citep{Rigby2017}, applying our current masking and source-plane measurement methods to J17-M2 yields a length of $\sim11\,\mathrm{kpc}$ for arc 1.1 from that model. This difference likely arises in part from the different masking methods and the greater depth of the JWST imaging, which allows us to detect more extended arc emission and therefore produces a larger mask than would be produced from the HST imaging. The difference in the size of the reconstructed source also reflects the different magnifications predicted by our model and J17-M2, with the inferred source-plane areas scaling consistently with the different magnifications.


\subsection{Implications for LEGGOS}

A central goal of LEGGOS is to resolve star formation in galaxies at cosmic noon on $\sim10-100$~pc scales, where individual star-forming regions begin to resemble giant molecular clouds. While JWST provides the necessary sensitivity and angular resolution, an important factor in achieving this goal is the accuracy of the lens model.

The improved constraint density in our model leads to a stable and well-localized magnification field, particularly in the vicinity of the primary arc where the scientific measurements are made. This directly reduces the model-dependent scatter in inferred clump properties, such as sizes and luminosities.

The magnification-uncertainty map in \autoref{fig:magnif} illustrates the posterior stability of our full model.  While the magnification is highest near the critical curve, the fractional uncertainty along the giant arc is among the lowest in \autoref{fig:magnif}, reaching $\sigma_\mu/|\mu_{\rm best}| \sim 0.03-0.09$ (statistical) $\pm 0.07$ (systematic). This represents a remarkably low level of model-to-model variation in a region that is typically expected to be highly sensitive to small perturbations in the lens model. The small uncertainty of the magnification along the arc reflects the dense sampling of the lensing potential provided by the large number of clumps. In contrast, the fractional uncertainty increases in regions with fewer constraints, particularly toward the western side of the field, highlighting the strong dependence of magnification precision on the local density of lensing constraints. The improved precision in clump magnifications, reflected by the low fractional uncertainties along the arc, reduces the lensing contribution to the statistical uncertainties in inferred luminosities and derived properties such as stellar mass and star-formation rate.

With lensing uncertainties reduced, LEGGOS can more robustly probe the physical nature of star-forming regions on unprecedented scales. This opens the door to connecting the small-scale physics of star formation and feedback to the global growth of galaxies during cosmic noon, enabling one of the central goals of the survey.

This also highlights the relative importance of the limiting factors for resolved studies of high-redshift galaxies: with JWST providing high signal-to-noise and angular resolution, systematic uncertainties in magnification and source-plane mapping become an increasingly important contributor to the overall uncertainty, alongside other modeling uncertainties such as those in stellar population and SED analyses. Continued improvements in lens modeling, such as spectroscopic confirmation of secondary systems, will therefore be critical for fully realizing the scientific potential of LEGGOS, as well as other studies of this lensing system.

\subsection{Limitations}

Despite the significant improvement in magnification uncertainties enabled by the new lensing analysis, several limitations remain. The absence of spectroscopic redshifts for all secondary arc systems introduces uncertainty in the mass distribution, especially away from the primary arc.

Degeneracies inherent to strong lensing allow multiple models to reproduce the observed image positions while yielding different magnification fields. As we discussed in \autoref{subsec:systematic_uncert} and \autoref{subsec:mass}, systematic uncertainties associated with lens modeling choices are not fully captured by the statistical errors reported here. While MCMC sampling explores the allowed parameter space within the adopted model, it does not account for alternative model choices or parameterizations that may also reproduce the observed constraints. This includes uncertainties in the outer mass profile, the treatment of cluster-member galaxies, the lack of spectroscopic redshifts for secondary systems, and the inclusion of additional lens components such as external shear. These effects are most pronounced in regions with sparse constraints and for quantities that depend sensitively on the magnification. Our comparison of independent \lenstool\ and \wslap\ models of RCS 0327 indicates that method-dependent magnification differences are modest at most clump positions, and may exceed 10\% in eight of the identified clumps: images 1 and 2 of clumps 1.1 and 1.2, and images 1 and 3 of clumps 1a.7 and 1a.8, which are $<2''$ from the critical curve.

Improvements to the lens model can be obtained by measuring and incorporating spectroscopic redshifts of the secondary arcs to better constrain the mass distribution, comparing reconstructions from independent lens modeling methods to directly assess systematic uncertainties due to lensing algorithms, and combining information from weak lensing to constrain the outskirts of the cluster and the large-scale structure around it.

\section{Conclusion}\label{sec:conclusion}

We have presented a new strong lensing model of \clustershort\ developed as part of the JWST LEGGOS survey, with the goal of enabling robust, spatially resolved studies of star formation in the highly magnified galaxy \galaxyshort\ at \galz. This work was motivated by the need for precise and internally consistent lens models to interpret JWST observations at $\sim10-100$~pc scales, as magnification and deflection uncertainties directly scale inferred physical properties.

Using deep JWST NIRCam imaging and NIRSpec spectroscopy, we identified an expanded set of multiple image constraints, including $>$20 clumps along the primary arc and spectroscopic confirmation that the edge structures belong to the same source. These constraints were incorporated into a fully parametric lens model constructed with \lenstool, including cluster-scale halos, galaxy-scale perturbations, and an external shear component. The model reproduces the observed image configuration with an RMS of $0\farcs11$ and provides well-constrained magnification and deflection maps tailored for JWST-based analyses.

We measure a total magnification of \magtot\ for \galaxyshort, consistent with previous work but with better constrained uncertainties. Along the giant arc we find fractional magnification uncertainties of $\sigma_\mu/|\mu_{\rm best}| \sim 0.03-0.09$, representing extremely low model-to-model variation in a region typically sensitive to model perturbations. For individual clump magnifications, we find a substantial reduction in the fractional uncertainty by a factor of $\sim 2.6-8.3\times$ compared to previous works, and estimate roughly $\sim7\%$ systematic uncertainty in these magnifications values, increasing for clumps located within $\sim2''$ of the critical curve. These results demonstrate how the increased density of constraints stabilizes the local lensing potential and magnification field, leading to more robust magnification estimates. As a result, the inferred properties of individual star-forming clumps, such as sizes and luminosities, are less sensitive to model-dependent variations, enabling more reliable interpretation of the intrinsic structure of the galaxy and the physical properties of its star-forming clumps.

The source-plane projections show consistent morphology across independent images, with a physical extent of $7.10^{+0.77}_{-1.51}$ kpc and multiple compact clumps. The agreement in morphology demonstrates that the lens model provides a robust mapping between the image- and source-plane in the region constrained by the arc.

These results highlight the opportunity to study the building blocks of star formation in high-redshift galaxies with unprecedented detail. With JWST providing high signal-to-noise imaging and strong lensing boosting the effective physical resolution to sub-kpc scales, it is now possible to probe internal structure in distant galaxies at a level that was previously inaccessible. Fully realizing this potential, however, requires accurate and precise lens models, since the inferred magnification and source-plane mapping directly shape the physical interpretation of these resolved structures.

Looking forward, further improvements to this model will primarily come from spectroscopic confirmation of secondary lensed systems already identified in the current data, which will provide tighter constraints on source redshifts and reduce degeneracies in the lens model. Additionally, combining strong and weak lensing analyses would better constrain the large-scale mass distribution and the influence of structures along the line of sight.  

The model presented here provides a robust lensing framework for future LEGGOS analyses of \galaxyshort, including detailed studies of star-forming clumps, stellar populations, and feedback processes at parsec scales. By combining strong gravitational lensing with JWST observations, LEGGOS is beginning to bridge the gap between the small-scale physics of star formation and the large-scale growth of galaxies during cosmic noon.

\vspace{5mm}
\begin{acknowledgements} 
We thank the anonymous referee for an insightful and constructive review and for their suggestions that greatly improved the paper.
This work is based on observations made with the NASA/ESA/CSA JWST. The data were obtained from the Mikulski Archive for Space Telescopes at the Space Telescope Science Institute, which is operated by the Association of Universities for Research in Astronomy, Inc., under NASA contract NAS 5-03127 for JWST. These observations are associated with JWST programs GO-03843 and GO-04125.

This work is based on observations made with the NASA/ESA Hubble Space Telescope obtained from the Space Telescope Science Institute, which is operated by the Association of Universities for Research in Astronomy, Inc., under NASA contract NAS 5–26555. These observations are associated with program HST GO-13003.

Support for JWST programs GO-03843 and GO-04125 was provided by NASA through a grant from the Space Telescope Science Institute, which is operated by the Association of Universities for Research in Astronomy, Inc., under NASA contract NAS 5-03127.

GK would like to thank the Baum Grant and Fellowship at the University of Washington for support during this work, as well as the ALMA Ambassador Program (administered by NAASC and NRAO). GK also thanks the International Space Science Institute (ISSI), Bern, for their hospitality, financial support and collaboration during the time of editing and commenting on this manuscript. 

This research was supported in part by the University of Pittsburgh Center for Research Computing, RRID:SCR\_022735, through the resources provided. Specifically, this work used the H2P/MPI cluster, which is supported by NSF award number OAC-2117681.

TER-T is supported by the Swedish Research Council grant 2022-04805.
\end{acknowledgements}

\facilities{JWST: NIRCam, NIRSpec; HST: WFC3}

\software{Astropy \citep{astropy:2013,astropy:2018,astropy:2022},
Matplotlib \citep{Hunter:2007},
SciPy \citep{2020SciPy-NMeth},
NumPy \citep{harris2020array}
\lenstool\ \citep{Jullo2007}
SAOImage DS9 \citep{ds9}
Source Extractor \citep{Bertin1996}
MATLAB \citep{matlab} 
MATLAB Astronomy and Astrophysics Toolbox \citep[MAAT][]{Ofek2014}, DrizzlePac \citep{driz},
Prospector \citep{Johnson_2021}}

\appendix
\section{Time Delay Prediction}\label{sec:appendix}

The so called ``Hubble tension'' that emerged in the last decade \cite[e.g.,][]{DiValentino21,Kamionkowski23,Tully23} re-motivated measuring $H_0$ from time delays between multiple images of strongly-lensed variable sources or transients. While a variable source has not yet been detected in \galaxyshort, its high star formation rate ($8.5^{+8.0}_{-0.4}\, M_{\odot}$ yr$^{-1}$, \citealt{Rigby2017}) could make it a prime target for monitoring for supernovae and other transient phenomena. In this Appendix, we provide a model-prediction of the time delays in \galaxyshort, on a clump-by-clump basis, to assist in planning of future observations, gravitational time delay statistics, and be immediately available for time delay analyses should a supernova is discovered anywhere in this arc in the future.  

The time delay between two images of the same source, $\Delta\tau_{ij}$, is calculated from the lensing potential through the Fermat potential \citep[e.g.,][]{Schneider85}: 
\begin{equation}\label{eq.fermat}
\tau(\vec\theta,\vec\beta) = \frac{1+z_l}{c}\frac{D_{l}D_{s}}{D_{ls}}\bigg[\frac{1}{2}(\vec\theta-\vec\beta)^2-\psi(\vec\theta)\bigg],
\end{equation}
\begin{equation}\label{eq.dt}
\Delta\tau_{ij} = \tau(\vec\theta_i,\vec\beta)-\tau(\vec\theta_j,\vec\beta),
\end{equation}
where $\vec\beta$ and $\vec\theta$ are the source and image-plane position, respectively; $z_l$ is the redshift of the lens, $D_{l}$, $D_{s}$ and $D_{ls}$ 
are the angular diameter distances to the lens, to the source, and from the lens to the source, respectively; 
and $\psi$ is the lensing potential. 
\autoref{tab:timedelay} lists the time delays between clumps in images 2 and 3 of \galaxyshort\ relative to image 1 of each identified clump. The time delay is calculated from the best-fit model, and an estimate of the statistical uncertainty is given as the standard variation of the distribution of time delays from the 300 models sampled from the MCMC. The time delay is measured at the observed location of each clump. We find that clumps in image 1 (the middle image) trail their counterparts in images 2 and 3 (the north and south images). Time delays range from $\sim$2-3 weeks between clump pairs that straddle the critical curves, to $\sim1.3$ years between clumps that are farthest from their counterparts.

\begin{longdeluxetable}{ccccc}
\tablecaption{Model-predicted time delays in \galaxyshort\ \label{tab:timedelay}}
\tablehead{
\colhead{Clump ID} &
\colhead{$\Delta\tau_{2-1}$ (days)} &
\colhead{$\sigma_{2-1}$ (days)} &
\colhead{$\Delta\tau_{3-1}$ (days)} &
\colhead{$\sigma_{3-1}$ (days)}
}
\startdata
 11 &      -13.1 &   1.0 &  \nodata & \nodata \\
 12 &      -23.3 &   1.5 &  \nodata & \nodata\\
 13 &      -45.2 &   2.9 &   -471.9 &  19.5\\
 14 &      -54.1 &   3.6 &  \nodata & \nodata\\
 15 &      -64.2 &   4.4 & -417.1 &  16.7 \\
 16 &      -77.3 &   5.3 & -389.5 &  15.2\\
 17 &      -95.3 &   5.7 & -405.3 &  15.0\\
 18 &      -96.8 &   6.8 & -350.3 &  13.2\\
 19 &      -121.4 &   8.5 & -308.5 &  11.2\\
110 &     -132.4 &   9.7 & -276.3 &  10.0\\
111 &      -141.0 &  10.4 & -251.6 &   9.0\\
112 &      -174.7 &  10.8 &  \nodata & \nodata\\
113 &      -189.8 &  12.4 &  \nodata & \nodata\\
114 &      -195.6 &  13.6 & -185.9 &   6.2\\
115 &      -239.8 &  14.3 & -185.7 &   5.8\\
116 &      -295.0 &  16.3 & -142.7 &   4.6\\
 1a1 &      -88.7 &   8.1 & -273.3 &  10.7\\
 1a2 &      -110.6 &   9.7 & -241.0 &   9.0\\
 1a3 &      -127.0 &  10.9 & -220.6 &   7.9\\
 1a4 &      -145.3 &  12.2 & -193.8 &   6.8\\
 1a5 &      -190.8 &  14.6 & -146.7 &   5.0\\
 1a6 &      -285.5 &  17.4 & -90.3 &   3.0\\
 1a7 &      -436.3 &   20.3 &  -33.4 &   1.5\\
 1a8 &      -455.9 &   20.8 &    -21.9 &   1.0\\
\enddata
\tablecomments{Time delay predictions for clumps the giant arc \galaxyshort. $\Delta\tau$ is the arrival time of each clump relative to image 1, measured in days from the best-fit model at the observed positions. $\sigma$ is the standard deviation of the distribution of time delays in the same 300 models sampled from the MCMC exploration of the parameter space.}
\end{longdeluxetable}

\newpage

\section{Photometric Measurements of Secondary Arcs}\label{sec:appendix_magnitdues}
In this appendix, we provide the photometry of secondary arcs used in the analysis, based on imaging in 4 HST filters and 10 JWST filters, tabulated in \autoref{tab:secondary_arc_photometry}.

\begin{deluxetable*}{cccccccc}
\tabletypesize{\scriptsize}
\tablecaption{Aperture Photometry of Secondary Arcs in \clustername
\label{tab:secondary_arc_photometry}}
\tablehead{
\colhead{Arc ID} &
\colhead{F390W} &
\colhead{F606W} &
\colhead{F070W} &
\colhead{F090W} &
\colhead{F105W} &
\colhead{F115W} &
\colhead{F150W}
}
\startdata
2.1 &
$26.29^{+0.60}_{-0.28}$ &
$25.91^{+0.17}_{-0.16}$ &
$26.42^{+0.23}_{-0.16}$ &
$26.33^{+0.26}_{-0.23}$ &
$26.18^{+0.42}_{-0.31}$ &
$25.66^{+0.17}_{-0.18}$ &
$25.52^{+0.18}_{-0.15}$ \\
2.2 &
$26.28^{+0.40}_{-0.24}$ &
$26.03^{+0.14}_{-0.11}$ &
$26.20^{+0.13}_{-0.09}$ &
$25.93^{+0.17}_{-0.11}$ &
$25.73^{+0.23}_{-0.15}$ &
$26.29^{+0.40}_{-0.20}$ &
$25.84^{+0.17}_{-0.15}$ \\
\hline
3.1 &
$25.42^{+0.19}_{-0.17}$ &
$24.31^{+0.03}_{-0.03}$ &
$24.43^{+0.04}_{-0.03}$ &
$24.29^{+0.07}_{-0.02}$ &
$24.17^{+0.06}_{-0.05}$ &
$24.19^{+0.04}_{-0.04}$ &
$23.83^{+0.03}_{-0.04}$ \\
3.2 &
$25.30^{+0.33}_{-0.19}$ &
$24.35^{+0.06}_{-0.05}$ &
$24.56^{+0.06}_{-0.05}$ &
$24.45^{+0.08}_{-0.07}$ &
$24.20^{+0.09}_{-0.07}$ &
$24.31^{+0.10}_{-0.07}$ &
$23.96^{+0.09}_{-0.05}$ \\
3.3 &
$27.34^{+1.96}_{-0.61}$ &
$25.48^{+0.07}_{-0.08}$ &
$25.55^{+0.07}_{-0.06}$ &
$25.42^{+0.08}_{-0.07}$ &
$25.34^{+0.12}_{-0.11}$ &
$25.27^{+0.09}_{-0.06}$ &
$24.82^{+0.07}_{-0.06}$ \\
3.4 &
$25.44^{+0.26}_{-0.20}$ &
$24.61^{+0.06}_{-0.05}$ &
$24.81^{+0.06}_{-0.05}$ &
$24.76^{+0.05}_{-0.07}$ &
$24.67^{+0.10}_{-0.11}$ &
$24.49^{+0.07}_{-0.08}$ &
$24.36^{+0.07}_{-0.07}$ \\
3.5 &
$26.98^{+1.25}_{-0.69}$ &
$25.52^{+0.11}_{-0.09}$ &
$25.45^{+0.14}_{-0.07}$ &
$25.52^{+0.13}_{-0.11}$ &
$25.34^{+0.19}_{-0.16}$ &
$25.19^{+0.14}_{-0.12}$ &
$24.73^{+0.13}_{-0.09}$ \\
\hline
4.1 &
$>28.04$ &
$26.98^{+0.29}_{-0.24}$ &
$26.51^{+0.11}_{-0.11}$ &
$26.16^{+0.08}_{-0.12}$ &
$26.49^{+0.35}_{-0.26}$ &
$26.28^{+0.16}_{-0.13}$ &
$26.28^{+0.19}_{-0.15}$ \\
4.2 &
$>28.18$ &
$27.26^{+0.33}_{-0.27}$ &
$26.51^{+0.10}_{-0.07}$ &
$26.39^{+0.14}_{-0.13}$ &
$26.89^{+0.61}_{-0.31}$ &
$26.31^{+0.16}_{-0.09}$ &
$26.24^{+0.14}_{-0.10}$ \\
\hline
\noalign{\vskip 6pt}
Arc ID & F160W & F182M & F200W & F277W & F356W & F444W & F480M \\
\hline
2.1 &
$25.48^{+0.37}_{-0.14}$ &
$25.71^{+0.34}_{-0.29}$ &
$25.75^{+0.34}_{-0.20}$ &
$25.64^{+0.15}_{-0.10}$ &
$25.50^{+0.14}_{-0.10}$ &
$25.55^{+0.13}_{-0.10}$ &
$25.38^{+0.33}_{-0.24}$ \\
2.2 &
$25.84^{+0.25}_{-0.19}$ &
$26.64^{+0.82}_{-0.31}$ &
$25.58^{+0.16}_{-0.11}$ &
$25.80^{+0.12}_{-0.08}$ &
$25.70^{+0.13}_{-0.07}$ &
$25.67^{+0.07}_{-0.06}$ &
$25.53^{+0.21}_{-0.17}$ \\
\hline
3.1 &
$23.92^{+0.06}_{-0.04}$ &
$23.19^{+0.03}_{-0.03}$ &
$23.99^{+0.05}_{-0.04}$ &
$24.02^{+0.04}_{-0.02}$ &
$23.90^{+0.02}_{-0.03}$ &
$24.01^{+0.03}_{-0.02}$ &
$24.05^{+0.04}_{-0.06}$ \\
3.2 &
$23.88^{+0.08}_{-0.06}$ &
$23.28^{+0.04}_{-0.07}$ &
$24.04^{+0.11}_{-0.07}$ &
$24.21^{+0.08}_{-0.05}$ &
$23.99^{+0.05}_{-0.04}$ &
$24.10^{+0.05}_{-0.04}$ &
$24.05^{+0.11}_{-0.07}$ \\
3.3 &
$24.98^{+0.10}_{-0.08}$ &
$24.15^{+0.05}_{-0.04}$ &
$24.92^{+0.09}_{-0.06}$ &
$24.99^{+0.04}_{-0.04}$ &
$24.81^{+0.04}_{-0.02}$ &
$24.87^{+0.03}_{-0.03}$ &
$24.76^{+0.08}_{-0.08}$ \\
3.4 &
$24.52^{+0.09}_{-0.09}$ &
$23.68^{+0.06}_{-0.05}$ &
$24.36^{+0.10}_{-0.07}$ &
$24.46^{+0.06}_{-0.03}$ &
$24.37^{+0.06}_{-0.04}$ &
$24.50^{+0.05}_{-0.05}$ &
$24.27^{+0.11}_{-0.07}$ \\
3.5 &
$25.11^{+0.18}_{-0.12}$ &
$24.29^{+0.12}_{-0.09}$ &
$24.48^{+0.08}_{-0.07}$ &
$24.54^{+0.08}_{-0.04}$ &
$25.04^{+0.12}_{-0.07}$ &
$25.28^{+0.10}_{-0.06}$ &
$24.66^{+0.15}_{-0.10}$ \\
\hline
4.1 &
$27.02^{+0.39}_{-0.31}$ &
$25.99^{+0.20}_{-0.16}$ &
$26.18^{+0.18}_{-0.10}$ &
$25.43^{+0.04}_{-0.04}$ &
$25.87^{+0.07}_{-0.05}$ &
$26.40^{+0.09}_{-0.11}$ &
$26.69^{+0.33}_{-0.25}$ \\
4.2 &
$26.62^{+0.28}_{-0.20}$ &
$26.28^{+0.29}_{-0.13}$ &
$26.28^{+0.15}_{-0.09}$ &
$25.32^{+0.02}_{-0.03}$ &
$25.96^{+0.09}_{-0.05}$ &
$26.55^{+0.07}_{-0.08}$ &
$26.40^{+0.19}_{-0.15}$ \\
\enddata
\tablecomments{Observed AB magnitudes measured using aperture photometry. The imaging was PSF-matched to the F480M data, and foreground light was modeled and subtracted where necessary. Magnitudes have not been corrected for lensing magnification. The F390W entries for arcs 4.1 and 4.2 are lower limits on magnitude, corresponding to upper limits on flux.}
\end{deluxetable*}

\bibliography{masterbib}{}
\bibliographystyle{aasjournal}

\end{document}